\documentclass[a4paper,11pt]{article}
\usepackage[usenames]{color}
\usepackage[normalem]{ulem}
\usepackage{amsfonts,amssymb}
\usepackage{theorem}
\usepackage{cancel}
\usepackage[dvips]{graphicx}

\def\g5{\gamma_{_\chi}}

\newcommand{\comment}[1]{}

\begin{document}
\comment
{
\par
\rightline{MIT, October 10,  2015 }
\rightline{Revised MIT, October 15,  2015 }
\par
}
\bf
\begin{center}{On effective Chern-Simons Term induced \\
by a Local CPT-Violating Coupling\\
 using {\huge $\gamma_5$} in  Dimensional Regularization
\footnote{\tt This work is supported in part by funds provided 
by the U.S. Department
of Energy (D.O.E.) under cooperative research agreement \#DE FG02-05ER41360}
}
\end{center}
\normalsize \rm
{
\large
\rm
\centerline{
Ruggero~Ferrari\footnote{e-mail: {\tt ruggferr@mit.edu}}$^{ac}$
and
Mario Raciti\footnote{e-mail: {\tt mario.raciti@mi.infn.it}}$^{bc}$}
\normalsize
\smallskip
\begin{center}
$^a$
Center for Theoretical Physics\\
Laboratory for Nuclear Science
and Department of Physics\\
Massachusetts Institute of Technology\\
Cambridge, Massachusetts 02139\\
$^b$Dipartimento di Fisica, Universit\`a degli Studi di Milano\\
$^c$
INFN, Sezione di Milano\\
via Celoria 16, I-20133 Milano, Italy

(MIT-CTP4680, IFUM-1041-FT, October 2015)
\end{center}
}   
\normalsize
\bf
\rm
\begin{quotation}{\bf Abstract: }
We resume  a long-standing, yet not forgotten, debate
on whether a Chern-Simons birefringence  can be generated
by a local term {$b_\mu\bar\psi\gamma^\mu \gamma_5\psi$ }
in the Lagrangian (where $b_\mu$ are constants).
\par
In the present paper we implement a new way
of managing  $\gamma_5$ in dimensional regularization.
Gauge invariance in the underlying theory (QED) is enforced
by this choice of defining divergent amplitudes.
\par
We investigate the singular behavior
of the vector meson two-point-function 
around the $m^2=0$ and $p^2=0$ point. We find that the coefficient
of the effective Chern-Simons can be finite or zero. 
It depends on how one takes the limits: they cannot
be interchanged due to the associate change of symmetry.
\par
For $m^2=0$ we evaluate also the self-mass of the photon at the second order
in  $b_\mu$. We find zero.

\end{quotation}
\newpage
\section{Introduction}
\label{sec:CPT}
There has been an interesting and long debate 
(see for instance \cite{Colladay:1996iz}-
\cite{Altschul:2004gs} )
on whether the Lorentz- and CPT-violating Chern-Simons 
{
effective action term
\begin{eqnarray}
\Delta {\cal S} = \int d^4 x\frac{1}{2} c^\mu \varepsilon_{\mu\alpha\beta\gamma}
F^{\alpha\beta} A^\gamma
\label{CPT.01}
\end{eqnarray}
might be generated by a local Lorentz- and CPT-violating
axial vector term in conventional QED}
{
\begin{eqnarray}
{\cal S}_{\rm extended} = \Lambda^{D-4}\int d^D x \Big[-\frac{1}{2}
\partial_\mu A_\nu \partial^\mu A^\nu
+\bar \psi\Big(i\not\!\partial-e \not\!\!A -m -\not\! b\gamma_5\Big)\psi
\Big],
\label{CPT.02}
\end{eqnarray}
where $b_\mu$ are constants and $\Lambda $ is the scale 
for dimensional regularization.
} 
\par
We discuss this problem at one loop in the perturbative
expansion. We use dimensional regularization extended
to $\gamma_5$,  as briefly
presented in the Section \ref{sec:DR}. The extension $\gamma_5$ 
to generic dimension $D$ has been proposed recently 
in Refs. \cite{Ferrari:2014jqa} and \cite{Ferrari:2015mha}.
The choice of dimensional regularization for the present problem is dictated by
the need of preserving the gauge invariance of the underlying QED
and of profiting of the advantages in actual computations
(e.g. the validity of formal properties of integration as:
shift in the integration variables and consistent use of divergent integrals).
\par
There have been previous attempts to use dimensional regularization
 \cite{Chung:1998jv}. 
Our results are in disagreement with the
cited paper. The origin of it might depend on the
use of different $\gamma_5$ extension, i.e. 't Hooft-Veltman's
and anticommuting $\gamma_5$.
We insist that our approach has many nice properties
as cyclicity of the gamma's trace and Lorentz covariance.
\par
Our results are in agreement with previous calculations where
QED gauge invariance is enforced through the regularization
procedure of Pauli-Villars.
\par  
In this framework the one-loop induced Chern-Simons with 
$m^2\not =0$ turns out
to vanish in the limit $p^2=0$ (on-shell photons). 
The result removes some ambiguities pointed out
in early works and never resolved by the use of dimensional
regularization because of the $\gamma_5$ problem.
\par
Our detailed analysis shows that the point $m^2=0$
$p^2=0$ is singular. The limit depends on the sequence in the
$(m^2,p^2)$ variables. 
\par 
We show that this peculiarity is due to a change in the
symmetry property of the model. If one puts $m^2=0$
right at the beginning the $b_\mu$ can be removed by
a local chiral transformation (at the classical level),
as noted by many authors. 
Thus only a local term survives (as in eq. (\ref{CPT.01})) similar
to the ABJ anomaly.
\par
We further investigate the properties of the $m^2=0, p^2\not =0$ case
by evaluating the photon two-point-function at second order
of $b_\mu$ in the one-loop approximation. We demonstrate that 
the amplitude must be gauge invariant. The explicit evaluation
shows that it is actually null. Thus in this case the quantum corrections
do not modify the independence from $b_\mu$ of the amplitudes,
which is present in the classical approximation. 
%
%
\section{Managing $\gamma_5$ in Dimensional Regularization}
\label{sec:DR}
While the extension of the gamma's to generic $D$ dimensions
is considered straightforward since the algebra
\begin{eqnarray}
\{\gamma_\mu,~\gamma_\nu\}=g_{\mu\nu}
\label{DR.1}
\end{eqnarray}
doesn't change with $D$, the situation with $\gamma_5$ is much
more complex. There is no proof that the extension to non-integer 
$D$ is possible and, if it is, we do not know how to do it.
To mark this important point we use $\gamma_5$  for $D=4$
and $\gamma_\chi$ for its extension (if any). Thus, for instance, 
we do not know what 
\begin{eqnarray}
\{\gamma_\chi,~\gamma_\nu\}
\label{DR.2}
\end{eqnarray}
might be. 
\par
The consequences of this fact are important in many respects.
One, often neglected, is that the invariance properties of
the action change when the gamma algebra is promoted to $D$
dimensions. For instance in $D=4$ dimensions the kinetic
term 
\begin{eqnarray}
\int d^4 x \bar \psi \gamma^\mu\partial_\mu \psi
\label{DR.3}
\end{eqnarray}
is invariant under global chiral transformations
\begin{eqnarray}
\psi \to e^{i\alpha\gamma_5}\psi.
\label{DR.4}
\end{eqnarray}
Therefore its Noether current is conserved (apart
possible anomalies). But in generic $D$ dimensions the 
action
\begin{eqnarray}
\int d^D x \bar \psi \gamma^\mu\partial_\mu \psi
\label{DR.5}
\end{eqnarray}
might \underline{not} be invariant under 
\begin{eqnarray}
\psi \to e^{i\alpha\gamma_\chi}\psi
\label{DR.6}
\end{eqnarray}
because the (anti)commutation relations (\ref{DR.2}) is
not necessarily null.
\par
The second point is the problem of evaluation of the trace, where
one or more $\gamma_\chi$ are present. To our experience the 
two difficulties are deeply intertwined. For instance
the anomaly of the axial current cancels the terms responsible
of the violation of the invariance of the generating functional. 
See \cite{Ferrari:2014jqa} Sec. 9 for a detailed discussion.
\par
We give a bird view of the approach implemented in the present
paper. Some details are not given here, but they can be 
traced in Refs.  \cite{Ferrari:2014jqa} and \cite{Ferrari:2015mha}.
\begin{enumerate}
\item It is assumed that an extension $\gamma_\chi$
exists so that a trace can be defined
\begin{eqnarray}
Tr(p)\equiv
Tr\Big (\gamma_\chi\not\!p_1\not\!p_2\not\!p_3\dots
\Big )
=Tr\Big (\gamma_\chi\gamma^\alpha \gamma^\beta  \gamma^\rho  \dots 
\Big ) p_{1\alpha} p_{2\beta}p_{3\rho}\dots
\label{DR.7}
\end{eqnarray}
The above expression can be generalized to many ${\gamma_\chi}'s$
in generic positions and the ${p\,}'s$ might be replaced by 
polarization vectors and by Lorentz 
covariant tensors, e.g. $g_{\alpha\rho}$ but not $\varepsilon_{\mu\nu\rho\sigma}$. 
It is important that none of the Lorentz indices are left free:
they appear only as dummy variables to sum over    { without specifying
their value (e.g. $v_\mu$ and $w_\mu$ are not provided, but $v^\mu w_\mu$
is an admissible quantity assuming real or complex values)}.
Lorentz covariance and cyclicity are required.
\item  In the neighborhood of $D=4$ the trace is assumed to
admit an expansion
\begin{eqnarray}
Tr(p)
= \sum _{h=0} A_h(p) (D-4)^h,\quad h \in {\mathcal N},
\label{DR.02}
\end{eqnarray}
where $A_h(p)$ are Lorentz invariants in $D=4$ dimensions (
the tensor $\varepsilon_{\mu\nu\rho\sigma}$ might be present).
\item
By assuming that the limit $D=4$ is smooth, one gets  
\begin{eqnarray}
\{\gamma_\chi,\gamma_\mu\} = {\cal O}(D-4),~ \forall \mu.
\label{DR.03}
\end{eqnarray}
\end{enumerate}
Clearly from step 1. to step 2. the Lorentz symmetry is
restricted to $D=4$.
\par
In the case of a single $\gamma_\chi$ a typical trick,
in order to evaluate the trace, is the following (cyclicity
is essential)
\begin{eqnarray}&&
Tr\Big(\gamma_\chi \not\!p_1\not\!p_2\dots \not\!p_k\Big)
\nonumber\\&&
=- 
Tr\Big( \not\!p_1  \gamma_\chi  \not\!p_2\dots \not\!p_k\Big)
+
Tr\Big(\gamma_\chi\Big\{ \not\!p_1,   \not\!p_2\dots \not\!p_k\Big\}\Big)
\label{DR.04}
\end{eqnarray}
The above relation is very intriguing. The first term in the RHS
yields the usual gamma's algebra in $D=4$, while the second
provides the ${\cal O}(D-4)$ term.
\par
To summarize, we assume  smooth dependence of traces
from $D$ around $D=4$, but at the same time we will avoid the evaluation
of any algebra where the (anti)commutation of $\gamma_\chi$ is involved. 
We might profit
of identities like (\ref{DR.04}). Lorentz covariance and cyclicity
are not negotiable. 
Details are in Refs. \cite{Ferrari:2014jqa} and \cite{Ferrari:2015mha},
where the ABJ anomaly and the isoscalar anomaly in $SU(2)$ nonabelian
chiral gauge theory are explicitly evaluated. In particular in the nonabelian
chiral gauge case, the many-$\gamma_\chi$ puzzle is solved.
\par
In the present paper for the calculation we use the template of Ref. 
\cite{Ferrari:2014jqa} with minor and straightforward changes.
{
\section{Peculiarities of the $m^2=0$ Case}
\label{sec:zerom}
In the classical approximation the $b_\mu$-term
can be removed by a local chiral transformation if $m^2=0$.
\par
The argument runs as follows 
\comment
{(the vector potential will be 
dropped later on for simplicity)}. The generating functional is obtained
by path integral on the \underline{classical} fields
\begin{eqnarray}&&
Z(J, \eta, \bar\eta ) = \int {\cal D}[A_\mu,\psi,\bar\psi]
\exp i\int d^4 x\Big(
\bar \psi\Big(i\not\!\partial-e \not\!\! A  -\not\! b\gamma_5\Big)\psi 
\nonumber\\&&
+J_\mu A^\mu+\bar\eta\psi+ \bar\psi \eta
\Big),
\label{zerom.1}
\end{eqnarray}
where $J_\mu, \eta, \bar \eta$ are external sources.
By performing a chiral transformation on the dummy
fields (classical and  in $D=4$)
\begin{eqnarray}
\psi \to e^{-ib_\mu x^\mu\gamma_5}\psi
\label{zerom.2}
\end{eqnarray}
we get
\begin{eqnarray}&&
Z(J, \eta, \bar\eta ) = \int {\cal D}[A_\mu,\psi,\bar\psi]
\exp i\int d^4 x\Big(
\bar \psi\Big(i\not\!\partial-e \not\!\! A  \Big)\psi 
\nonumber\\&&
+J_\mu A^\mu+\bar\eta e^{-ib_\mu x^\mu\gamma_5}\psi+ \bar\psi e^{-ib_\mu x^\mu\gamma_5} \eta
\Big)
\nonumber\\&&
=Z_{(0)}(J, e^{-ib_\mu x^\mu\gamma_5}\eta, \bar\eta e^{-ib_\mu x^\mu\gamma_5}),
\label{zerom.3}
\end{eqnarray}
where $Z_{(0)}$ indicates that the action  has no $b_\mu$ term.
\par
The result in eq. (\ref{zerom.3}) is very strong: if one considers
only $A_\mu$-amplitudes there is no dependence from  $b_\mu$.
Some dependence from  $b_\mu$ might emerge from quantum
corrections as we will see later on.
\par
Before dropping the subject, it is amusing to see how the original
action is restored by starting from the identity in
(\ref{zerom.3}). For simplicity we drop $A_\mu$.
The integration over the Fermi fields yields the usual result
\begin{eqnarray}&&
Z_{(0)}(\eta, \bar\eta ) = 
\exp \Big(\int d^4x d^4y \bar\eta(x) \frac{i}{(2\pi)^4}\int d^4p
\frac{e^{-ip(x-y)}}{\not\! p +i\epsilon}\eta(y)\Big).
\label{zerom.4}
\end{eqnarray}
According to eq. (\ref{zerom.3}) we have to replace
$\eta$ by $e^{-ib_\mu x^\mu\gamma_5}\eta$. Use
\begin{eqnarray}
e^{-ib_\mu x^\mu\gamma_5} = e^{-ib_\mu x^\mu} \gamma_++
e^{ib_\mu x^\mu}\gamma_-,  \qquad \gamma_\pm \equiv  \frac{1\pm\gamma_5}{2}
\label{zerom.5}
\end{eqnarray}
and get
\begin{eqnarray}&&
Z_{(0)}(e^{-ib_\mu x^\mu\gamma_5}\eta, \bar\eta e^{-ib_\mu x^\mu\gamma_5})
=
\exp \Big(\int d^4x d^4y \bar\eta(x) 
\nonumber\\&&
\frac{i}{(2\pi)^4}\int d^4p
e^{-ip(x-y)}\Big[\frac{1}{\not\! p-\not\!b +i\epsilon}
\gamma_-+
\frac{1}{\not\! p+\not\!b +i\epsilon}
\gamma_+\Big]
\eta(y)\Big)
\label{zerom.6}
\end{eqnarray}
With a little algebra the relations can be written in the form
\begin{eqnarray}&&
\frac{1}{\not\! p-\not\!b }
\gamma_-
=
\Big[\Big(1-\not\! b \frac{1}{\not\! p}\Big)\not\!p\Big]^{-1}\gamma_-
=\frac{1}{\not\! p}\Big(\sum_{k=0}^\infty 
\left[\not\! b \frac{1}{\not\! p}\right]^k\Big)\gamma_-
\nonumber\\&&
\frac{1}{\not\! p+\not\!b }
\gamma_+
=
\Big[\Big(1+\not\! b \frac{1}{\not\! p}\Big)\not\!p\Big]^{-1}\gamma_+
=\frac{1}{\not\! p}\Big(\sum_{k=0}^\infty 
\left[-\not\! b \frac{1}{\not\! p}\right]^k\Big)\gamma_+
\label{zerom.7}
\end{eqnarray}
Finally the sum yields
\begin{eqnarray}&&
\frac{1}{\not\! p-\not\!b }
\gamma_- + 
\frac{1}{\not\! p+\not\!b }
\gamma_+
=\frac{1}{\not\! p}\Big(\sum_{k=0}^\infty 
\left[-\not\! b \frac{1}{\not\! p}\gamma_5\right]^k\Big)
\label{zerom.8a}
\\&&
=\frac{1}{\not\! p}
\Big(1+\not\! b \frac{1}{\not\! p}\gamma_5\Big)^{-1}
=\Big[\Big(1+\not\! b \frac{1}{\not\! p}\gamma_5\Big)\not\! p\Big]^{-1}
=\Big(\not\! p-\not\! b \gamma_5\Big)^{-1}
\label{zerom.8b}
\end{eqnarray}
i.e. we are back again to the action (\ref{zerom.1}) (by restoring $A_\mu$).
\par
The above example suggests that the correct expansion is  in number of 
loops. At each order in the number of loops the amplitude must
be $b_\mu-$independent, when properly summed over all powers of $b_\mu$.
This conclusion is unavoidable. 
Thus the only possibility in order to have a $b_\mu-$ dependence
of the two-point-function of the vector mesons is by means
of an anomaly of the axial current. More precisely: one uses
the expansion in powers of $b_\mu$ as in equations (\ref{zerom.8a})
and (\ref{zerom.8b})
for the fermion propagators of the $A_\mu$ two-point-function.
Among the numerous graphs the triangular ones are expected to be
anomalous and therefore a $b_\mu$ dependence arises.
This will be evaluated  in the Section \ref{sec:one}. 
\par
This is an unexpected situation. The parameter $b_\mu$ cannot
be fixed phenomenologically by some S-matrix measurement. 
If the anomaly is measured, we get out the value of $b_\mu$,
a fact that is not very much instructive, since the interacting
term with fermions disappears   {(we could use an action
with no $b_\mu$).}
\par\noindent
{\bf Comment:} In the process of quantization of the field
theory (\ref{zerom.1}) one needs a regularization procedure 
in order to deal with ill-defined amplitudes. We use dimensional
regularization and consequently the algebra involving
$\gamma_\chi$ is far from trivial (see Section \ref{sec:DR}). 
We envisage that classically equivalent Feynman rules
yield different results at the quantum level, since there is
no possibility to enforce normalization conditions ($b_\mu$
disappears from the action at the classical level) to force the
theories at the same results.   {We could use a perturbative expansion
where the unperturbed  propagator is the conventional free  Fermi-Dirac
as in eq. 
(\ref{zerom.3}) or its power series expansion in $b_\mu$ (\ref{zerom.8a})
or the {\sl non perturbative} approach (\ref{zerom.8b}).
We choose the power series expansion in $b_\mu$ because we do not know
how to deal with $\gamma_\chi$ in the denominator in presence of
dimensional regularization. Moreover this choice
allows a comparison with the existing literature.}  
}
\section{One-loop Calculations}
\label{sec:one}
In this work we give the detailed calculations at one loop 
of the ensuing terms, due to the 
$b_\mu$ external source given in eq. (\ref{CPT.02}).
At the first order we expect a Chern-Simons term,
while at the second order a self-mass of the photon 
might arise.
\par
The result of the computations is quite surprising, thus
we present the whole matter systematically.   {The scenario 
is the following.}
\par
At first order in $b_\mu$ and for  $p^2\not =0,m^2=0$ the 
Chern-Simons term is non zero ($c_\mu\not =0$ in eq. (\ref{CPT.01})).
The calculation is very close to the derivation of
Adler-Bardeen-Jackiw Anomaly \cite{Ferrari:2014jqa}.
\par
Instead zero is the value of $c_\mu$ for $p^2 =0,m^2\not =0$.
\par
For  $p^2<< m^2\not =0$ the coefficient $c_\mu\sim \frac{p^2}{m^2}$.
Thus there is a discontinuity at  $(p^2 =0, m^2=0)$.
\par
At the second order in $b_\mu$ we evaluate the photon
self-energy at $m^2=0$. We demonstrate that Ward identity
is satisfied. The explicit  computation
yields an indisputable result: zero.
\par
It should be stressed that the use of dimensional regularization
is freeing us from problems present in other regularization schemes:
there is no ambiguities on shifting by a constant the
integration variable of the inner momentum of the loop.
\par
We deal with the $\gamma_\chi$ by following the rules outlined
in Section \ref{sec:DR}. We bring together all factors
of the $D-4$ pole, due to the inner momentum integration.
We verify that a common $D-4$ factor appears to cancel the pole,
to yield a final finite result. Only this part of the calculation
requires  care in the use of   $\gamma_\chi$. The rest of the
calculation can be performed at $D=4$, since every term is
finite.
\par
We start from the formulae of Section 8 of the paper 
\cite{Ferrari:2014jqa}. Eventually we take the appropriate limits.

\section{ABJ Anomaly ($p^2\not=0, m^2=0$)}
\label{sec:ano}
We consider a massive fermion triangle, where one vertex
is given by an axial current. We will remove the
mass when necessary. Thus we consider the integral
($p$ is the incoming momentum on the vertex $\sigma$ and $k$ on $\rho$; crossed
graph will be considered at the end)
\begin{eqnarray}&&
J_{\mu\rho\sigma}(k,p)={-i}  {\Lambda^{4-D}}
\int \frac{d^D q}{(2\pi)^D}~
\frac{1}
{
[(q-k)^2-m^2][ q^2-m^2][ (q+p)^2-m^2]
}
\nonumber\\&&
 Tr ~\Bigl\{\gamma_\mu\gamma_\chi~
[(q-k)^\alpha\gamma_\alpha+m]  ~\gamma_\rho ~[q^\beta\gamma_\beta~ +m]\gamma_\sigma~
[(q+p)^\iota\gamma_\iota+m]
\Bigr\}
\label{ano.1}
\end{eqnarray}
Now we use Feynman parameterization
and get
\begin{eqnarray}&&
= {-i}  2 \Lambda^{4-D}
\int_0^1 dx\int_0^x dy \int \frac{d^D q}{(2\pi)^D}~
\Big\{(q+px-py-ky)^2-(px-py-ky)^2
\nonumber\\&&
 -m^2 +p^2(x-y)+k^2y\Big\}^{-3}
\nonumber\\&& 
Tr ~\Bigl\{\gamma_\mu\gamma_\chi~
[(q-k)^\alpha\gamma_\alpha+m]  ~\gamma_\rho ~[q^\beta\gamma_\beta~ +m]\gamma_\sigma~
[(q+p)^\iota\gamma_\iota+m]
\Bigr\}.
\label{jackiw.2}
\end{eqnarray}
We change variable
\begin{eqnarray}
q \to q+r, \qquad
{
r \equiv yk-xp+yp}
\label{ano.2}
\end{eqnarray}
and implement symmetric integration
\begin{eqnarray}&&
 ={-i}  2 \Lambda^{4-D}
\int_0^1 dx\int_0^x dy \int \frac{d^D q}{(2\pi)^D}~
\Big\{q^2-(px-py-ky)^2
\nonumber\\&&
 -m^2 +p^2(x-y)+k^2y\Big\}^{-3}
\nonumber\\&& 
Tr ~\Bigl\{\gamma_\mu\gamma_\chi~
[(q+r-k)^\alpha\gamma_\alpha+m]  ~\gamma_\rho ~[(q+r)^\beta
\gamma_\beta~ +m]\gamma_\sigma~
\nonumber\\&& 
[(q+r+p)^\iota\gamma_\iota+m]
\Bigr\}.
\label{jackiw.3}
\end{eqnarray}
%
{
Let us define
\begin{eqnarray}
\Delta = m^2 -p^2(x-y)-k^2y +(px-py-ky)^2
\label{jackiw.4}
\end{eqnarray}
Thus
\begin{eqnarray}&&
 ={-i}  2  { \Lambda^{4-D}}
\int_0^1 dx\int_0^x dy \int \frac{d^D q}{(2\pi)^D}~
\Big\{q^2    - \Delta \Big\}^{-3}
\nonumber\\&& 
Tr ~\Bigl\{\gamma_\mu\gamma_\chi~
[(q+yk-xp+yp-k)^\alpha\gamma_\alpha+m]  
\nonumber\\&& 
\gamma_\rho ~[(q+yk-xp+yp)^\beta
\gamma_\beta~ +m]\gamma_\sigma~
\nonumber\\&& 
[(q+yk-xp+yp+p)^\iota\gamma_\iota+m]
\Bigr\}.
\label{jackiw.5}
\end{eqnarray}
}
After symmetric integration over $q$ we can split the integral
into a divergent part (the terms proportional to $m$ are associated
to a $Tr$ with an odd number of gamma's and one $\gamma_\chi$; this
is expected to be zero for $D\sim 4$)
\begin{eqnarray}&&
J_{\mu\rho\sigma}^{\rm DIV}(k,p)=
{-i}
 \frac{1}{2}
\int_0^1  dx \int_0^x dy
Tr ~\Bigl\{\gamma_\mu\gamma_\chi
\gamma_\alpha\gamma_\rho 
\gamma_\beta \gamma_\sigma\gamma_\iota \Bigr\}
\nonumber\\&& 
 \Bigl((yk-xp+yp+p)^\iota g^{\alpha\beta}
 +(yk-xp+yp)^\beta g^{\alpha\iota}
+(yk-xp+yp-k)^\alpha g^{\beta\iota}\Bigr)
\nonumber\\&&
\Big(-\frac{i}{(4\pi)^2} \Big)\Big[\frac{2}{D-4}+\gamma +2
-\ln 4\pi+\ln(\Delta) \Big]
  {
\Big(1- \frac{D-4}{4}- (D-4)\ln \Lambda\Big)}
\nonumber\\&&
\label{ano.5}
\end{eqnarray}
and finite part
\begin{eqnarray}&&
J_{\mu\rho\sigma}^{\rm CONV}(k,p)=
{-i}
 2
\int_0^1 dx\int_0^x dy 
 \frac{-i}{2(4\pi)^2} 
\frac{1}{{\Delta}}
\nonumber\\&&
Tr ~\Bigl\{\gamma_\mu\gamma_\chi~
[(yk-xp+yp-k)^\alpha\gamma_\alpha+m]  
~\gamma_\rho ~[(yk-xp+yp)^\beta
\gamma_\beta~ +m]\gamma_\sigma~
\nonumber\\&& 
[(yk-xp+yp+p)^\iota\gamma_\iota+m]
\Bigr\}
.
\label{ano.6}
\end{eqnarray}
In front of the two amplitudes (\ref{ano.5}) and (\ref{ano.6})
the gamma's trace must be expanded in powers of $(D-4)$
as required by eq. (\ref{ano.1}). For the finite part in eq.
(\ref{ano.6}) we can use the $D=4$ expression, but for the
divergent part one needs also the linear part in $(D-4)$.
\par
For $D\to 4$ we get
\begin{eqnarray}&&
\int \frac{d^D q}{(2\pi)^D} \frac{q^2}{(q^2-m^2)^3}
\nonumber\\&&
= -\frac{i}{(4\pi)^2} \Big [ \frac{2}{D-4}+\gamma
+\frac{1}{2} + \ln\big(\frac{m^2}{4\pi}\big)\Big] + {\cal O}(D-4)
\label{ano.5.0.old}
\end{eqnarray}
and 
\begin{eqnarray}
\int \frac{d^D q}{(2\pi)^D} \frac{1}{(q^2-m^2)^3}
= -\frac{i}{2(4\pi)^2} 
\frac{1}{m^2}   + {\cal O}(D-4).
\label{ano.6.0.old}
\end{eqnarray}
The splitting in divergent and convergent parts (\ref{ano.5}) and
(\ref{ano.6}) is arbitrary. It is done only in order to make
the exposition simpler.
\subsection{Calculation of Divergent Part ($p^2\not=0,m^2 =0$)}
Now we consider to the limit $m^2=0$.
Since we look at the effective Chern-Simons term
we take the conditions
\begin{eqnarray}
k=-p.
\label{zerom.10}
\end{eqnarray}
The divergent part (\ref{ano.5}) becomes
\begin{eqnarray}&&
J_{\mu\rho\sigma}^{\rm DIV}(p) =
{-i}
 \frac{1}{2}
\int_0^1  dx \int_0^x dy
Tr ~\Bigl\{\gamma_\mu\gamma_\chi
\gamma_\alpha\gamma_\rho 
\gamma_\beta \gamma_\sigma\gamma_\iota \Bigr\}
\nonumber\\&& 
 \Bigl((-xp+p)^\iota  g^{\alpha\beta}
 +(-xp)^\beta   g^{\alpha\iota}
+(-xp+p)^\alpha  g^{\beta\iota}\Bigr)
\nonumber\\&&
\Big(-\frac{i}{(4\pi)^2} \Big)\Big[\frac{2}{D-4}   {
- \frac{1}{2}- 2 \ln \Lambda}
\nonumber\\&&
+\gamma +2
-\ln 4\pi+\ln( -p^2(x-x^2) ) \Big]
\label{zerom.11}
\end{eqnarray}
The trace is evaluated according to the rules of Ref.
\cite{Ferrari:2014jqa} Sec.8.
\begin{eqnarray}&&
{-i}
 \frac{1}{2}
\int_0^1  dx \int_0^x dy
Tr ~\Bigl\{\gamma_\mu\gamma_\chi\gamma_\rho 
 \gamma_\sigma\gamma_\iota \Bigr\}p^\iota
\nonumber\\&& 
 \Bigl((1-x)(2-D)
 +x (6-D)
+(1-x)(2-D)\Bigr)
\nonumber\\&&
\Big(-\frac{i}{(4\pi)^2} \Big)\Big[\frac{2}{D-4}  {
- \frac{1}{2}- 2 \ln \Lambda}
\nonumber\\&&
+\gamma +2
-\ln 4\pi+\ln( -p^2(x-x^2) ) \Big]
\nonumber\\&& 
=
{-i}
 \frac{1}{2}
\int_0^1  dx x
Tr ~\Bigl\{\gamma_\mu\gamma_\chi\gamma_\rho 
 \gamma_\sigma\gamma_\iota \Bigr\}p^\iota
\nonumber\\&& 
 \Bigl(6x -4 + (D-4)(x-2)
 \Bigr)
\nonumber\\&&
\Big(-\frac{i}{(4\pi)^2} \Big)\Big[\frac{2}{D-4}   {
- \frac{1}{2}- 2 \ln \Lambda}
\nonumber\\&&
+\gamma +2
-\ln 4\pi+\ln( -p^2(x-x^2) ) \Big]
\label{zerom.12}
\end{eqnarray}
Since $\int_0^1  dx x(6x-4)=0$ the $x$-independent part inside
the square brackets gives zero
contribution. What is left is
\begin{eqnarray}&&
=
{-i}\Big(-\frac{i}{(4\pi)^2} \Big)
 \frac{1}{2}
Tr ~\Bigl\{\gamma_\mu\gamma_\chi\gamma_\rho 
 \gamma_\sigma\gamma_\iota \Bigr\}p^\iota
\nonumber\\&& 
 \Bigl(\int_0^1  dx (6x^2 -4x)\ln(x-x^2) - \frac{4}{3}
 \Bigr)
\nonumber\\&&
\label{zerom.13}
\end{eqnarray}
Elementary integration yields 
\begin{eqnarray}
\int_0^1 dx (6x^2-4x) \ln x(1-x) = -\frac{1}{3}
\label{zerom.16}
\end{eqnarray}
Finally eq. (\ref{zerom.13}) becomes
\begin{eqnarray}&&
J_{\mu\rho\sigma}^{\rm DIV}(p) 
=
\frac{1}{(4\pi)^2} 
 \frac{1}{2}
Tr ~\Bigl\{\gamma_\mu\gamma_\chi\gamma_\rho 
 \gamma_\sigma\gamma_\iota \Bigr\}p^\iota
 \frac{5}{3}
\nonumber\\&&
\label{zerom.17}
\end{eqnarray}
  {
Notice that the scale parameter $\Lambda$
has dropped out from the final result.}
\subsection{Calculation of convergent Part ($p^2\not=0,m^2 =0$)}
\label{sec:conv}
Now we evaluate the convergent part (\ref{ano.6}) in the 
particular kinematic condition (\ref{zerom.10}). Every integral
is finite; therefore the algebra is in  $D=4$
\begin{eqnarray}&&
J_{\mu\rho\sigma}^{\rm CONV}(p) =
{-i}
 2
\int_0^1 dx x
 \frac{-i}{2(4\pi)^2} 
\frac{1}{-p^2(x-x^2)}
\nonumber\\&&
Tr ~\Bigl\{\gamma_\mu\gamma_\chi~
[(-xp+p)^\alpha\gamma_\alpha]  
~\gamma_\rho ~[(-xp)^\beta
\gamma_\beta~ ]\gamma_\sigma~
\nonumber\\&& 
[(-xp+p)^\iota\gamma_\iota]
\Bigr\}
\nonumber\\&&
=
\int_0^1 dx x
 \frac{1}{(4\pi)^2} 
\frac{1}{p^2(x-x^2)}
\nonumber\\&&
Tr ~\Bigl\{\gamma_\mu\gamma_\chi~ p^\alpha\gamma_\alpha
(-x+1) ~\gamma_\rho ~(-x)p^\beta\gamma_\beta~ \gamma_\sigma~
\nonumber\\&& 
(-x+1)p^\iota\gamma_\iota
\Bigr\}
\nonumber\\&&
=
\int_0^1 dx x
 \frac{1}{(4\pi)^2} 
\frac{1}{p^2(x-x^2)}
\nonumber\\&&
Tr ~\Bigl\{\gamma_\mu\gamma_\chi~ p^2
(-x+1) ~\gamma_\rho ~x  \gamma_\sigma~
(-x+1)p^\iota\gamma_\iota
\Bigr\}
\nonumber\\&&
=
\int_0^1 dx x
 \frac{1}{(4\pi)^2} (1-x)
Tr ~\Bigl\{\gamma_\mu\gamma_\chi 
 \gamma_\rho   \gamma_\sigma~
p^\iota\gamma_\iota
\Bigr\}
\nonumber\\&&
= \frac{1}{(4\pi)^2} \frac{1}{6}Tr ~\Bigl\{\gamma_\mu\gamma_\chi 
 \gamma_\rho   \gamma_\sigma
\gamma_\iota
\Bigr\}p^\iota
.
\label{conv.1}
\end{eqnarray}
\subsection{Final Result for $m^2=0$}
We add the divergent and finite parts in eqs.(\ref{zerom.17}) and (\ref{conv.1})
and their crossed terms
\begin{eqnarray}&&
J_{\mu\rho\sigma}^{\rm DIV}(p)+
J_{\mu\rho\sigma}^{\rm CONV}(p)+
J_{\mu\sigma\rho}^{\rm DIV}(-p)+
J_{\mu\sigma\rho}^{\rm CONV}(-p)
\nonumber\\&& 
=2\frac{1}{(4\pi)^2} Tr ~\Bigl\{\gamma_\mu\gamma_\chi
\gamma_\rho  \gamma_\sigma
\gamma_\iota
\Bigr\}p^\iota
.
\label{total.1}
\end{eqnarray}
This is consistent with the ABJ anomaly of the axial current, i.e.
one gets the current from the $k\not = -p$ anomaly and then takes
the limit value $k = -p$.
\par   {
Thus we conclude that  the CPT-violating term in the action 
(\ref{CPT.02}) induces a Cern-Simons effective  amplitude
(\ref{CPT.01}) with
\begin{eqnarray}
c_\mu = \frac{b_\mu}{4 \pi^2},
\label{total.2}
\end{eqnarray}
}
{
by taking $Tr\{\gamma_5 \gamma_\mu\gamma_\nu\gamma_\rho\gamma_\sigma\}
=i4 \varepsilon_{\mu\nu\rho\sigma}$.}
\par
The result in eq. (\ref{total.2}) has been discussed at length in the
literature. See for instance Refs. 
\cite{Chung:1999pt} and \cite{PerezVictoria:1999uh}.

\section{Induced Chern-Simons for $p^2=0,m^2\not =0$}
\label{sec:self}
For comparison with the Section \ref{sec:ano} and for
phenomenological applications it is convenient now to 
discuss the case  $p^2=0,m^2\not =0$.

We consider the divergent part (\ref{ano.5}) at $p^2=0$
\begin{eqnarray}&&
T_{\mu\rho\sigma}^{\rm DIV}=
{-i}
 \frac{2}{D}
Tr ~\Bigl\{\gamma_\mu\gamma_\chi
\gamma_\alpha\gamma_\rho 
\gamma_\beta \gamma_\sigma\gamma_\iota \Bigr\}
 \Bigl(\frac{1}{6}  g^{\alpha\beta}
p^\iota -\frac{1}{3}  g^{\alpha\iota}p^\beta
+\frac{1}{6}  g^{\beta\iota}p^\alpha\Bigr)
\nonumber\\&&
\Big(-\frac{i}{(4\pi)^2} \Big)\Big[\frac{2}{D-4}+\gamma +2
-\ln 4\pi+\ln(m^2) \Big]
\nonumber\\&&
=
{-i}
 \frac{2}{D}
 \Bigl(\frac{2-D}{6}Tr ~\Bigl\{\gamma_\mu\gamma_\chi
\gamma_\rho  \gamma_\sigma\gamma_\iota \Bigr\}
p^\iota
\nonumber\\&&
 -\frac{6-D}{3}Tr ~\Bigl\{\gamma_\mu\gamma_\chi
\gamma_\rho 
\gamma_\beta \gamma_\sigma \Bigr\}p^\beta
+\frac{2-D}{6}
Tr ~\Bigl\{\gamma_\mu\gamma_\chi
\gamma_\alpha\gamma_\rho 
 \gamma_\sigma \Bigr\}p^\alpha\Bigr)
\nonumber\\&&
\Big(-\frac{i}{(4\pi)^2} \Big)\Big[\frac{2}{D-4}+\gamma +2
-\ln 4\pi+\ln(m^2) \Big]
\nonumber\\&&
=
{-i}
 \frac{2}{D}\frac{2}{3}(4-D)Tr ~\Bigl\{\gamma_\mu\gamma_\chi
\gamma_\rho  \gamma_\sigma\gamma_\iota \Bigr\}
p^\iota
\nonumber\\&&
\Big(-\frac{i}{(4\pi)^2} \Big)\Big[\frac{2}{D-4}+\gamma +2
-\ln 4\pi+\ln(m^2) \Big]
\nonumber\\&&
=
 \frac{2}{3}
\frac{1}{(4\pi)^2}Tr ~\Bigl\{\gamma_\mu\gamma_\chi
\gamma_\rho  \gamma_\sigma\gamma_\iota \Bigr\}
p^\iota
\label{ano.5.1}
\end{eqnarray}
%
\subsection{Calculation of Convergent Part $p^2=0,m^2\not =0$.}
Thus eq. (\ref{ano.6}) becomes (at $p^2=0$)
\begin{eqnarray}&&
T_{\mu\rho\sigma}^{\rm CONV}
={-i}
 2
\int_0^1 xdx (-1)\frac{i}{2(4\pi)^2} 
\frac{1}{m^2}
\nonumber\\&&
m^2Tr ~\Bigl\{\gamma_\mu\gamma_\chi\gamma_\rho\gamma_\sigma
(1-x)p^\iota\gamma_\iota
+\gamma_\mu\gamma_\chi
  \gamma_\rho (-1)xp^\beta
\gamma_\beta\gamma_\sigma~
\nonumber\\&& 
+\gamma_\mu\gamma_\chi
(1-x)p^\alpha\gamma_\alpha  \gamma_\rho \gamma_\sigma
\Bigr\}
\nonumber\\&& 
=
-\frac{1}{(4\pi)^2} \int_0^1 xdxTr ~\Bigl\{
\gamma_\mu\gamma_\chi\gamma_\rho\gamma_\sigma\gamma_\iota\Bigr\} p^\iota
\Big(1-x +x +1-x   \Big)
\nonumber\\&& 
=
-\frac{1}{(4\pi)^2} \int_0^1 dx(2x-x^2)Tr ~\Bigl\{
\gamma_\mu\gamma_\chi\gamma_\rho\gamma_\sigma\gamma_\iota\Bigr\} p^\iota
\nonumber\\&& 
= -\frac{1}{(4\pi)^2}  (1-\frac{1}{3})Tr ~\Bigl\{
\gamma_\mu\gamma_\chi\gamma_\rho\gamma_\sigma\gamma_\iota\Bigr\} p^\iota
\nonumber\\&& 
=- \frac{2}{3}\frac{1}{(4\pi)^2}Tr ~\Bigl\{
\gamma_\mu\gamma_\chi\gamma_\rho\gamma_\sigma\gamma_\iota\Bigr\} p^\iota
.
\label{ano.7}
\end{eqnarray}
The crossed graph gives a factor 2
\begin{eqnarray}
T_{\mu\rho\sigma}^{\rm CONV+CROSSED}
=- \frac{4}{3}\frac{1}{(4\pi)^2}Tr ~\Bigl\{
\gamma_\mu\gamma_\chi\gamma_\rho\gamma_\sigma\gamma_\iota\Bigr\} p^\iota
.
\label{ano.70}
\end{eqnarray}
We add the divergent part eq. (\ref{ano.5.1}) (notice the 
difference with Section 8.1 of \cite{Ferrari:2014jqa} 
with $k=-p$) and the crossed divergent part 
\begin{eqnarray}
T_{\mu\rho\sigma}^{\rm DIV+CROSSED}
=\frac{{1}}{(4\pi)^2} \frac{4}{3}
Tr \Bigl(\gamma_\mu\gamma_\chi\gamma_\rho
\gamma_\sigma\gamma_\iota \Bigr) p^\iota.
\label{ano.12}
\end{eqnarray}
Finally one gets
\begin{eqnarray}
T_{\mu\rho\sigma}^{\rm CONV+CROSSED}+T_{\mu\rho\sigma}^{\rm DIV+CROSSED}
= 0.
\label{ano.13}
\end{eqnarray}
The result is surprisingly different from the one obtained
in Section \ref{sec:ano} in eq. (\ref{total.1}). The difference
between $m^2=0$ and $m^2\not =0$ has been discussed also
in Ref. \cite{PerezVictoria:2001ej}.
\par
Thus we investigate the dependence ($c_\mu$) from $p^2$ in a
massive theory.
\section{Photon Two-point Function at $|p^2|<<m^2 $}
\label{sec:virtual}
For virtual photons it is relevant to evaluate the
two-point function off-shell. However we hold the condition
\begin{eqnarray}
|p^2|<<m^2.
\label{virtual.1}
\end{eqnarray}
%
{
The general case is discussed in Appendix \ref{sec:gen}.
\par
For the present case we resume  eqs. (\ref{ano.5}) and (\ref{ano.6}) and expand
in $\frac{p^2}{m^2}$. The divergent part yields}
\begin{eqnarray}
&&
P_{\mu\rho\sigma}^{\rm DIV}=
{-i}
 \frac{2}{D}
\int_0^1 x dx
Tr ~\Bigl\{\gamma_\mu\gamma_\chi
\gamma_\alpha\gamma_\rho 
\gamma_\beta \gamma_\sigma\gamma_\iota \Bigr\}
\nonumber\\&& 
 \Bigl((1-x)  g^{\alpha\beta}
p^\iota -x  g^{\alpha\iota}p^\beta
+(1-x)  g^{\beta\iota}p^\alpha\Bigr)
\nonumber\\&&
\Big(-\frac{i}{(4\pi)^2} \Big)\Big[\frac{2}{D-4}+\gamma +2
-\ln 4\pi+\ln(m^2)- \frac{p^2}{m^2}(x-x^2) \Big]
\label{virtual.5}
\end{eqnarray}
and finite part
\begin{eqnarray}&&
P_{\mu\rho\sigma}^{\rm CONV}={-i}
 2
\int_0^1 xdx \frac{-i}{2(4\pi)^2}
\frac{1}{m^2-p^2(x-x^2)}
\nonumber\\&&
Tr ~\Bigl\{\gamma_\mu\gamma_\chi~
[(1-x)p^\alpha\gamma_\alpha+m]  ~\gamma_\rho ~[-xp^\beta
\gamma_\beta~ +m]\gamma_\sigma~
\nonumber\\&& 
[(1-x)p^\iota\gamma_\iota+m]
\Bigr\}
\nonumber\\&&
= - \frac{1}{(4\pi)^2}
\int_0^1 xdx \Big(
\frac{1}{m^2}+ (x-x^2)\frac{p^2}{m^4}\Big)
\nonumber\\&&
Tr ~\Bigl\{\gamma_\mu\gamma_\chi~
[(1-x)p^\alpha\gamma_\alpha+m]  ~\gamma_\rho ~[-xp^\beta
\gamma_\beta~ +m]\gamma_\sigma~
\nonumber\\&& 
[(1-x)p^\iota\gamma_\iota+m]
\Bigr\}
.
\label{virtual.6}
\end{eqnarray}
\subsection{Photon Two-point Function at $|p^2|<<m^2 $: Divergent Part}
%
We elaborate eq. (\ref{virtual.5}). We keep only the $p^2$
dependence
\begin{eqnarray}&&
P_{\mu\rho\sigma}^{\rm DIV}=
\frac{1}{(4\pi)^2}
 \frac{2}{D}\frac{p^2}{m^2}
Tr ~\Bigl\{\gamma_\mu\gamma_\chi
\gamma_\alpha\gamma_\rho 
\gamma_\beta \gamma_\sigma\gamma_\iota \Bigr\}
\nonumber\\&& 
\int_0^1 x dx
 \Bigl((1-x)  g^{\alpha\beta}
p^\iota -x  g^{\alpha\iota}p^\beta
+(1-x)  g^{\beta\iota}p^\alpha\Bigr)
(x-x^2) 
\nonumber\\&&
= 
\frac{1}{(4\pi)^2}
 \frac{2}{D}\frac{p^2}{m^2}
Tr ~\Bigl\{\gamma_\mu\gamma_\chi
\gamma_\alpha\gamma_\rho 
\gamma_\beta \gamma_\sigma\gamma_\iota \Bigr\}
\nonumber\\&& 
\int_0^1  dx
 \Bigl((x^2-2x^3+x^4)  g^{\alpha\beta}
p^\iota -(x^3-x^4)  g^{\alpha\iota}p^\beta
+(x^2-2x^3+x^4)  g^{\beta\iota}p^\alpha\Bigr)
\nonumber\\&&
= 
\frac{1}{(4\pi)^2}
 \frac{2}{D}\frac{p^2}{m^2}
Tr ~\Bigl\{\gamma_\mu\gamma_\chi
\gamma_\alpha\gamma_\rho 
\gamma_\beta \gamma_\sigma\gamma_\iota \Bigr\}
\nonumber\\&& 
 \Bigl((\frac{1}{3}-\frac{1}{2}+\frac{1}{5})  g^{\alpha\beta}
p^\iota -\frac{1}{20}  g^{\alpha\iota}p^\beta
+\frac{1}{30}  g^{\beta\iota}p^\alpha\Bigr)
\nonumber\\&& 
=\frac{1}{10}
\frac{1}{(4\pi)^2}
 \frac{2}{D}\frac{p^2}{m^2}
Tr ~\Bigl\{\gamma_\mu\gamma_\chi
\gamma_\alpha\gamma_\rho 
\gamma_\beta \gamma_\sigma\gamma_\iota \Bigr\}
\nonumber\\&& 
 \Bigl(\frac{1}{3}  g^{\alpha\beta}
p^\iota -\frac{1}{2}  g^{\alpha\iota}p^\beta
+\frac{1}{3}  g^{\beta\iota}p^\alpha\Bigr)
\nonumber\\&& 
=\frac{1}{10}
\frac{1}{(4\pi)^2}
 \frac{2}{D}\frac{p^2}{m^2}
\nonumber\\&& 
 \Bigl(\frac{2}{3}(2-D)Tr ~\Bigl\{\gamma_\mu\gamma_\chi
\gamma_\rho  \gamma_\sigma\gamma_\iota \Bigr\}
p^\iota -\frac{1}{2}(6-D)
Tr ~\Bigl\{\gamma_\mu\gamma_\chi
\gamma_\rho 
\gamma_\beta \gamma_\sigma \Bigr\}p^\beta\Bigr)
\nonumber\\&& 
=\frac{1}{10}
\frac{1}{(4\pi)^2}
 \frac{2}{D}\frac{p^2}{m^2}Tr ~\Bigl\{\gamma_\mu\gamma_\chi
\gamma_\rho  \gamma_\sigma\gamma_\iota \Bigr\}p^\iota 
 \Bigl(-\frac{4}{3}
+ 1\Bigr)
\nonumber\\&& 
=-\frac{1}{60}
\frac{1}{(4\pi)^2}
\frac{p^2}{m^2}Tr ~\Bigl\{\gamma_\mu\gamma_\chi
\gamma_\rho  \gamma_\sigma\gamma_\iota \Bigr\}p^\iota 
\label{virtual.7}
\end{eqnarray}
\subsection{Photon Two-point Function at $|p^2|<<m^2 $: Convergent Part}
In the finite part (\ref{virtual.6}) we have  two contributions: one from
$p^3$ and one from $m^2$.
\begin{eqnarray}&&
P_{\mu\rho\sigma}^{\rm CONV}=
 -
\int_0^1 xdx \frac{1}{(4\pi)^2} \Big(
\frac{1}{m^2}+ (x-x^2)\frac{p^2}{m^4}\Big)
\nonumber\\&&
Tr ~\Bigl\{\gamma_\mu\gamma_\chi~
[(1-x)p^\alpha\gamma_\alpha+m]  ~\gamma_\rho ~[-xp^\beta
\gamma_\beta~ +m]\gamma_\sigma~
\nonumber\\&& 
[(1-x)p^\iota\gamma_\iota+m]
\Bigr\}
\nonumber\\&&
= - \frac{1}{(4\pi)^2}
\int_0^1 xdx \Bigg(-x(1-x)^2
\frac{1}{m^2}Tr ~\Bigl\{\gamma_\mu\gamma_\chi~
p^\alpha\gamma_\alpha  ~\gamma_\rho ~p^\beta
\gamma_\beta~ \gamma_\sigma~
p^\iota\gamma_\iota
\Bigr\}
\nonumber\\&& 
+(x-x^2)\frac{p^2}{m^2} ~Tr ~\Bigl\{\gamma_\mu\gamma_\chi~\Big[
(1-x)\gamma_\rho\gamma_\sigma p^\iota\gamma_\iota
- x\gamma_\rho p^\beta\gamma_\beta\gamma_\sigma
+ (1-x)p^\alpha\gamma_\alpha\gamma_\rho\gamma_\sigma\Big]\Bigr\}
\Bigg)
\nonumber\\&&
= - \frac{1}{(4\pi)^2}\frac{p^2}{m^2}Tr ~\Bigl\{\gamma_\mu\gamma_\chi
\gamma_\rho\gamma_\sigma p^\iota\gamma_\iota\Big\}
\int_0^1 dx\Bigg[x^2(1-x)^2+ x^2(1-x)\Big((1-x)
+ x + (1-x)\Big)\Bigg]
\nonumber\\&&
= - \frac{1}{(4\pi)^2}\frac{p^2}{m^2}Tr ~\Bigl\{\gamma_\mu\gamma_\chi
\gamma_\rho\gamma_\sigma p^\iota\gamma_\iota\Big\}
\int_0^1 dx  (x^2-2x^3+x^4+2x^2-3x^3+x^4)
\nonumber\\&&
= - \frac{1}{(4\pi)^2}\frac{p^2}{m^2}Tr ~\Bigl\{\gamma_\mu\gamma_\chi
\gamma_\rho\gamma_\sigma p^\iota\gamma_\iota\Big\}
 (1-\frac{5}{4}+\frac{2}{5})
\nonumber\\&&
= - \frac{1}{(4\pi)^2}\frac{p^2}{m^2}Tr ~\Bigl\{\gamma_\mu\gamma_\chi
\gamma_\rho\gamma_\sigma p^\iota\gamma_\iota\Big\}
 \frac{20-25+8}{20}
\nonumber\\&&
= - \frac{1}{(4\pi)^2}\frac{p^2}{m^2}Tr ~\Bigl\{\gamma_\mu\gamma_\chi
\gamma_\rho\gamma_\sigma p^\iota\gamma_\iota\Big\}
 \frac{3}{20}
.
\label{virtual.8}
\end{eqnarray}
\subsection{Total}
%
\begin{eqnarray}
P_{\mu\rho\sigma}^{\rm DIV}+
P_{\mu\rho\sigma}^{\rm CONV}=
 - \frac{1}{(4\pi)^2}\frac{p^2}{m^2}Tr ~\Bigl\{\gamma_\mu\gamma_\chi
\gamma_\rho\gamma_\sigma p^\iota\gamma_\iota\Big\}
 \frac{10}{60}
\label{virtual.9}
\end{eqnarray}
Now we add the crossed term (multiply by 2)
\begin{eqnarray}
P_{\mu\rho\sigma}^{\rm DIV+CROSSED}+
P_{\mu\rho\sigma}^{\rm CONV+CROSSED}=
 - \frac{1}{(4\pi)^2}\frac{p^2}{m^2}Tr ~\Bigl\{\gamma_\mu\gamma_\chi
\gamma_\rho\gamma_\sigma p^\iota\gamma_\iota\Big\}
 \frac{1}{3}.
\label{virtual.10}
\end{eqnarray}
The result in eq. (\ref{virtual.10}) shows that there is continuity for $p^2 \to 0$
with the value (\ref{ano.13}) of Section \ref{sec:self}.

%
\section{Second Order in $b_\mu$: Self-Energy with $m^2=0$. }
\label{sec:box}
The calculation in Section \ref{sec:ano} shows that 
a finite result emerge from the conspiring of pole and 
zero in $D-4$; although the $b_\mu$ term in the action
can be removed by a chiral transformation as in eq. (\ref{zerom.3}). 
\par
It is natural to look for other cases where the Feynman integrals
provide poles in $(D-4)$. Thus
we consider the  photon two-point-function at the second order in $b_\mu$. 
Here we have three graphs.
One ($\prod^{(1)}_{\rho\sigma}$) where $\not\! b$ alternates with $\not\!\!\! A$
on the fermion line 
and  ($\prod^{_{(2)}}_{\rho\sigma}$) and ($\prod^{(3)}_{\rho\sigma}$) where
$\not\! b$ are consecutive. 
\par
Transversality is important issue since the amplitude is a
photon self-energy.
\subsection{Gauge Invariance at 1-loop: the Box ($m^2=0$).}
This important question must be discussed in detail. 
\par
One expects no problems since the Ward
identity should not be affected by the $CPT$-violating
term. I.e. the generating functional should obey the
transversality condition.  The presence of $\gamma_\chi$
is irrelevant for the derivation of the functional identity.
\par
It is however intriguing to see it in an elementary
derivation, i.e. by using
\begin{eqnarray}
&&
p^\rho
\frac{1}
{ [\not\!q +i\varepsilon ]}\gamma_\rho 
\frac{1}
{
[(\not\!q+\not\!p)+i\varepsilon]}
= \frac{1}
{ [\not\!q +i\varepsilon ]}-
\frac{1}
{
[(\not\!q+\not\!p)+i\varepsilon]}
\label{box.-1}
\end{eqnarray}

Let us consider the box diagrams one by one. 
\begin{eqnarray}&&
\prod_{\rho\sigma}(p)^{{(1)}}
={-}
\int \frac{d^D q}{(2\pi)^D}~ Tr ~\Bigl\{
\frac{1}
{ [\not\!q +i\varepsilon ]}\not \! b\gamma_\chi~
\frac{1}{\not \! q  +i\varepsilon} 
\gamma_\rho
\frac{1}
{
[(\not\!q+\not\!p)+i\varepsilon]}
\nonumber\\&&~
\not \! b\gamma_\chi~ 
\frac{1}{
[ \not \!q + \not\! p+i\varepsilon]
}
~\gamma_\sigma
\Bigr\}
\label{box.1}
\end{eqnarray}
Then to check gauge invariance we evaluate
\begin{eqnarray}&&
p^\rho\prod_{\rho\sigma}(p)^{(1)}
={-}
\int \frac{d^D q}{(2\pi)^D}~ Tr ~\Bigl\{
\frac{1}{ [\not\!q +i\varepsilon ]}
\not \! b\gamma_\chi~\frac{1}{\not \! q  +i\varepsilon} ~
\nonumber\\&&
\not \! b\gamma_\chi~ 
\frac{1}{[ \not \!q + \not\! p+i\varepsilon]}
~\gamma_\sigma
\nonumber\\&&
-
\frac{1}
{ [\not\!q +i\varepsilon ]}\not \! b\gamma_\chi~
\frac{1}
{
[(\not\!q+\not\!p)+i\varepsilon]}\not \! b\gamma_\chi~ 
\frac{1}{
[ \not \!q + \not\! p+i\varepsilon]
}
~\gamma_\sigma
\Bigr\}
\label{box.2}
\end{eqnarray}
%
%
Now we consider the graphs where both $b_\mu$ insertions are
on the same Fermionic line
\begin{eqnarray}&&
\prod_{\rho\sigma}(p)^{(2)}
={-}
\int \frac{d^D q}{(2\pi)^D}~ Tr ~\Bigl\{
~\gamma_\sigma
\frac{1}{\not \! q  +i\varepsilon} ~
\gamma_\rho
\frac{1}{
[(\not\!q+\not\!p)+i\varepsilon]}\not \! b\gamma_\chi~ 
\frac{1}{
[ \not \!q + \not\! p+i\varepsilon]
}\not \! b\gamma_\chi~
\nonumber\\&&
\frac{1}
{ [\not\!q+ \not\! p +i\varepsilon ]}
\Bigr\}
\nonumber\\&&
\label{box.1.1}
\end{eqnarray}
Then the divergence gives
\begin{eqnarray}&&
p^\rho
\prod_{\rho\sigma}(p)^{(2)}
={-}
\int \frac{d^D q}{(2\pi)^D}~ Tr ~\Bigl\{
~\gamma_\sigma
\frac{1}{\not \! q  +i\varepsilon} ~
\not \! b\gamma_\chi~ 
\frac{1}{
[ \not \!q + \not\! p+i\varepsilon]
}\not \! b\gamma_\chi~
\nonumber\\&&
\frac{1}
{ [\not\!q+ \not\! p +i\varepsilon ]}
\nonumber\\&&
-
~\gamma_\sigma
\frac{1}
{
[ \not \!q + \not\! p+i\varepsilon]
}\not \! b\gamma_\chi~
\frac{1}
{ [\not\!q+ \not\! p +i\varepsilon ]}
\not \! b\gamma_\chi~
\frac{1}
{ [\not\!q+ \not\! p +i\varepsilon ]}
\Bigr\}.
\label{box.2.1}
\end{eqnarray}
Finally the third graph (obtained from $\prod_{\rho\sigma}(p)^{(2)}$ by
$\rho \leftrightarrow \sigma, p\to -p$) gives
\begin{eqnarray}&&
\prod_{\rho\sigma}(p)^{(3)}
={-}
\int \frac{d^D q}{(2\pi)^D}~ Tr ~\Bigl\{\gamma_\rho
\frac{1}{\not \! q  +i\varepsilon} ~\gamma_\sigma
\frac{1}
{
[(\not\!q-\not\!p)+i\varepsilon]}\not \! b\gamma_\chi~ 
\nonumber\\&&
\frac{1}{
[ \not \!q - \not\! p+i\varepsilon]
}\not \! b\gamma_\chi~
\frac{1}
{ [\not\!q- \not\! p +i\varepsilon ]}
\Bigr\}
\label{box.4.1}
\end{eqnarray}
and its divergence
\begin{eqnarray}&&
p^\rho\prod_{\rho\sigma}(p)^{(3)}
={-}
\int \frac{d^D q}{(2\pi)^D}~ Tr ~\Bigl\{ ~\gamma_\sigma
\frac{1}
{
[(\not\!q-\not\!p)+i\varepsilon]}\not \! b\gamma_\chi~ 
\nonumber\\&&
\frac{1}{
[ \not \!q - \not\! p+i\varepsilon]
}\not \! b\gamma_\chi~
\frac{1}
{ [\not\!q- \not\! p +i\varepsilon ]}
\nonumber\\&&
-
\frac{1}{\not \! q  +i\varepsilon} ~\gamma_\sigma
\frac{1}
{
[(\not\!q-\not\!p)+i\varepsilon]}\not \! b\gamma_\chi~ 
\frac{1}{
[ \not \!q - \not\! p+i\varepsilon]
}\not \! b\gamma_\chi~
\Bigr\}
\label{box.5.1}
\end{eqnarray}
The equation above (\ref{box.2}), (\ref{box.2.1}) and (\ref{box.5.1}) show the
exact cancellation of the divergence of the two-point-function
\begin{eqnarray}
p^\rho\Big (\prod_{\rho\sigma}(p)^{(1)}+   
\prod_{\rho\sigma}(p)^{(2)}+
\prod_{\rho\sigma}(p)^{(3)} 
\Big)  =0
\label{box.6.1}
\end{eqnarray}
In particular the second in eq. (\ref{box.2.1}) cancels the first
in eq. (\ref{box.5.1}) through a change of variable. The first in 
(\ref{box.2}) cancels the second in (\ref{box.5.1}). Finally
the second in (\ref{box.2}) cancels the first in (\ref{box.2.1}).
 No $\gamma_\chi$ was involved in the proof. 
\par
However there is a flaw in the proof: the quantities present 
in the eqs. (\ref{box.2}), (\ref{box.2.1}) and (\ref{box.5.1}) 
are ill-defined due to ultraviolet divergences: the algebra of
$\gamma_\chi$ is needed for generic $D$. The verifications 
of the equations will be possible after a precise statement
about the algebra.
\par
 Only when we group all the $D-4$ poles together we
obtain a well-defined quantity. At that moment we can verify the
Ward identity.
\par
Transversality then requires the following general form
\begin{eqnarray}&&
\prod_{\rho\sigma}(p) = a_1 ~b^2~\Big[  p_\rho p_\sigma - p^2   g_{\rho \sigma}   \Big]
\nonumber\\&&
+\frac{a_2}{p^2} \Big[p^2 b_\rho b_\sigma + (pb)^2   g_{\rho\sigma}
- (pb)(p_\rho b_\sigma + b_\rho p_\sigma)
\Big].
\label{trivia.2.1}
\end{eqnarray}
%

%
%
%
\subsection{Divergent Parts: the Box $m^2=0$.}
%
We look at the divergent parts (by this we
denote the amplitude where  a $q^4$ power appears in
the numerator). Thus we do not consider the difference 
between eq. (\ref{bbox.1.10}) and  (\ref{bbox.1.11}).
Later on we will take into account also the $p^2$
dependence.

From eq. (\ref{box.1}) we have the
gamma's factor
\begin{eqnarray}&&
{\cal V}_{\rho\sigma}^{(1){\rm DIV}} = Tr\Big\{\gamma_\rho \not\! q \not\! b \gamma_\chi
\not \! q \gamma_\sigma \not \! q \not\! b \gamma_\chi \not\! q 
\Big\}
\nonumber\\&&
 = Tr\Big\{(-q^2\gamma_\rho +2 q_\rho\not\! q) \not\! b \gamma_\chi
(-q^2\gamma_\sigma +2 q_\sigma\not\! q) 
\not\! b \gamma_\chi 
\Big\}
\nonumber\\&&
 = q^4Tr\Big\{\gamma_\rho \not\! b \gamma_\chi
\gamma_\sigma \not\! b \gamma_\chi
\Big\}
- 2 q_\rho q^2Tr\Big\{\not\! q \not\! b \gamma_\chi
\gamma_\sigma \not\! b \gamma_\chi 
\Big\}
\nonumber\\&&
- 2 q_\sigma q^2Tr\Big\{\gamma_\rho \not\! b \gamma_\chi
  \not\! q \not\! b \gamma_\chi
\Big\}
+ 4q_\rho q_\sigma Tr\Big\{  \not\! q  \not\! b \gamma_\chi
  \not\! q \not\! b \gamma_\chi
\Big\}
\label{bbox.2}
\end{eqnarray}
The symmetric integration gives
\begin{eqnarray}&&
{\cal V}^{(1){\rm DIV}}_{\rho\sigma}
 = q^4Tr\Big\{\gamma_\rho \not\! b \gamma_\chi
\gamma_\sigma \not\! b \gamma_\chi
\Big\}
\nonumber\\&&
- \frac{2}{D}  q^4Tr\Big\{\gamma_\rho \not\! b \gamma_\chi
\gamma_\sigma \not\! b \gamma_\chi 
\Big\}
%
- \frac{2}{D}  q^4Tr\Big\{\gamma_\rho \not\! b \gamma_\chi
\gamma_\sigma \not\! b \gamma_\chi 
\Big\}
\nonumber\\&&
+ \frac{4q^4}{D(D+2)} \Big (
     g_{\rho\sigma}Tr\Big\{ \gamma_\tau \not\! b \gamma_\chi
  \gamma^\tau \not\! b \gamma_\chi\Big\}
+2Tr\Big\{ \gamma_\rho \not\! b \gamma_\chi
  \gamma_\sigma \not\! b \gamma_\chi\Big\}
\Big)
\nonumber\\&&
=q^4Tr\Big\{\gamma_\rho \not\! b \gamma_\chi
\gamma_\sigma \not\! b \gamma_\chi
\Big\}\Big(1-\frac{4}{D} + \frac{8}{D(D+2)}
\Big) 
\nonumber\\&&
+ \frac{4q^4}{D(D+2)} 
     g_{\rho\sigma}Tr\Big\{ \gamma_\tau \not\! b \gamma_\chi
  \gamma^\tau \not\! b \gamma_\chi\Big\}
\nonumber\\&&
=\frac{(D - 2)q^4}{D+2}Tr\Big\{\gamma_\rho \not\! b \gamma_\chi
\gamma_\sigma \not\! b \gamma_\chi
\Big\}
%
+ \frac{4q^4     g_{\rho\sigma}}{D(D+2)} 
Tr\Big\{ \gamma_\tau \not\! b \gamma_\chi
  \gamma^\tau \not\! b \gamma_\chi\Big\}
\label{bbox.3}
\end{eqnarray}
The other graphs yield
\begin{eqnarray}&&
{\cal V}_{\rho\sigma}^{(2){\rm DIV}} = Tr\Big\{ \not\! q \gamma_\rho \not\! q \not\! b \gamma_\chi
\not \! q \not\! b \gamma_\chi\not \! q \gamma_\sigma 
\Big\}
\nonumber\\&&
= 
Tr\Big\{ \Big[2 q_\rho \not \! q - q^2 \gamma_\rho \Big] \not\! b \gamma_\chi
\not \! q \not\! b \gamma_\chi\not \! q \gamma_\sigma 
\Big\}
\label{bbox.4.1}
\end{eqnarray}
By symmetric integration
\begin{eqnarray}&&
{\cal V}_{\rho\sigma}^{(2){\rm DIV}} 
=
\frac{2q^4}{D(D+2)} 
Tr\Big\{ \gamma_\rho\not\! b \gamma_\chi
\gamma_\tau \not\! b \gamma_\chi\gamma^\tau \gamma_\sigma 
%
+(2-D)\gamma_\sigma \not\! b \gamma_\chi\gamma_\rho\not\! b \gamma_\chi
\nonumber\\&&
+ \gamma_\tau \not\! b \gamma_\chi \gamma^\tau\not\! b \gamma_\chi\gamma_\rho\gamma_\sigma
\Big\}
- \frac{q^4}{D}\Big\{\gamma_\rho\not\! b \gamma_\chi \gamma_\tau\not\! b \gamma_\chi 
\gamma^\tau\gamma_\sigma
\Big\}
\nonumber\\&&
= -\frac{q^4}{D+2}Tr\Big\{ \not\! b \gamma_\chi
\gamma_\tau \not\! b \gamma_\chi\gamma^\tau \gamma_\sigma\gamma_\rho \Big\}
+
\frac{2q^4(2-D)}{D(D+2)} 
Tr\Big\{\gamma_\sigma \not\! b \gamma_\chi\gamma_\rho\not\! b \gamma_\chi  \Big\}
\nonumber\\&&
+\frac{2q^4}{D(D+2)} Tr\Big\{\gamma_\tau \not\! b \gamma_\chi \gamma^\tau\not\! b \gamma_\chi\gamma_\rho\gamma_\sigma
\Big\}
\label{bbox.4.2}
\end{eqnarray}
We add the third graph ($\rho \leftrightarrow \sigma $)
\begin{eqnarray}&&
{\cal V}_{\rho\sigma}^{(2){\rm DIV}} +
{\cal V}_{\rho\sigma}^{(3){\rm DIV}} 
\nonumber\\&&
= -\frac{2q^4}{D+2}     g_{\rho\sigma}Tr\Big\{ \not\! b \gamma_\chi
\gamma_\tau \not\! b \gamma_\chi\gamma^\tau  \Big\}
\nonumber\\&&
+
\frac{4q^4(2-D)}{D(D+2)} 
Tr\Big\{\gamma_\sigma \not\! b \gamma_\chi\gamma_\rho\not\! b \gamma_\chi  \Big\}
\nonumber\\&&
+\frac{4q^4}{D(D+2)}      g_{\rho\sigma}Tr\Big\{\gamma_\tau \not\! b \gamma_\chi \gamma^\tau\not\! b \gamma_\chi
\Big\}
\nonumber\\&&
=\frac{2q^4(2-D)}{D(D+2)} Tr\Big\{     g_{\rho\sigma} \not\! b \gamma_\chi
\gamma_\tau \not\! b \gamma_\chi\gamma^\tau 
+ 2 \gamma_\sigma \not\! b \gamma_\chi\gamma_\rho\not\! b \gamma_\chi 
\Big\}
\label{bbox.4.3}
\end{eqnarray}

\subsection{Pole Part: the Box $m^2=0$.}
Now we take the sum
\begin{eqnarray}&&
{\cal V}_{\rho\sigma}^{{\rm DIV}} 
=
{\cal V}_{\rho\sigma}^{(1){\rm DIV}} +
{\cal V}_{\rho\sigma}^{(2){\rm DIV}} +
{\cal V}_{\rho\sigma}^{(3){\rm DIV}} 
\nonumber\\&&
=- 2q^4     g_{\rho\sigma} \frac{(D-4)}{D(D+2)} Tr\Big\{\gamma_\tau
 \not\! b \gamma_\chi\gamma^\tau \not\! b \gamma_\chi
\Big\}
\nonumber\\&&
+ q^4\frac{(D-4)(D-2)}{D(D+2)} Tr\Big\{
\gamma_\sigma \not\! b \gamma_\chi\gamma_\rho\not\! b \gamma_\chi 
\Big\}
\label{bbox.6}
\end{eqnarray}
It vanishes for $D=4$; then we can use the elementary algebra
\footnote{A factor $ Tr \{{\mathcal I}\}$ will be neglected
throughout the paper.}
\begin{eqnarray}&&
{\cal V}_{\rho\sigma}^{\rm DIV} 
=2q^4     g_{\rho\sigma} b^2 
\frac{(D-4)(D-2)}{D(D+2)}
\nonumber\\&&
+ q^4  \frac{(D-2)}{D(D+2)}(D-4)\Big(2b_\rho b_\sigma -b^2     g_{\rho\sigma} 
\Big)
\nonumber\\&&
=q^4 \frac{(D-4)(D-2)}{D(D+2)}
\Big(2b_\rho b_\sigma+b^2     g_{\rho\sigma}
\Big)
\label{bbox.7}
\end{eqnarray}
Finally we get the pole part in $D-4$, form eqs. (\ref{box.1}), (\ref{box.1.1}), 
 (\ref{box.4.1}), (\ref{bbox.7}), (\ref{bbox.1.10}), (\ref{bbox.1.11}) and
(\ref{ano.5.0.2})
\begin{eqnarray}&&
\prod_{\rho\sigma}^{\rm POLE}= \frac{i}{(4\pi)^2} 
\frac{1}{6}
\Big(2b_\rho b_\sigma+b^2     g_{\rho\sigma}
\Big)
\label{bbox.8}
\end{eqnarray}
%
\subsection{Log Parts: the Box $m^2=0$.}
For the log parts we need the integrals in (\ref{bbox.1.10}) and (\ref{bbox.1.11})
The graph n. 1 yields the following divergent-log contribution.
From eq. (\ref{bbox.1.10}), (\ref{bbox.3}) (\ref{ano.5.0.2}) and (\ref{bbox.9})
we have
\begin{eqnarray}&&
{\prod_{\rho\sigma}^{\rm LOG}}^{_{(1)}}
=\frac{1}{3} \Big(Tr\Big\{\gamma_\rho \not\!b \gamma_\sigma\not\!b\Big\}
-     g_{\rho\sigma}Tr\Big\{\not\!b\not\!b\Big\} \Big)
\nonumber\\&&
- 6\int_0^1 dx\int_0^x dy  y (-\frac{i}{(4\pi)^2}) \ln(y(1-y))
\nonumber\\&&
=\frac{2}{3}
(b_\rho b_\sigma- b^2       g_{\rho\sigma})
(-\frac{5}{3})(\frac{i}{(4\pi)^2})
\nonumber\\&&
= -\frac{i}{(4\pi)^2}\frac{10}{9}
(b_\rho b_\sigma- b^2       g_{\rho\sigma})
\label{bbox.15}
\end{eqnarray}
For the graph n.2 we have similarly the following contributions.
Eqs. (\ref{bbox.1.11}), (\ref{bbox.4.2}), (\ref{ano.5.0.2}) and 
(\ref{bbox.10.1})  are  needed
\begin{eqnarray}&&
{\prod_{\rho\sigma}^{\rm LOG}}^{_{(2)}}
= - \frac{1}{6} 
\Big(Tr\Big\{\gamma_\rho \not\!b \gamma_\sigma\not\!b\Big\}
-     g_{\rho\sigma}Tr\Big\{\not\!b\not\!b\Big\} \Big)
\nonumber\\&&
 - 6\int_0^1 dx \frac{x^2}{2}   (-\frac{i}{(4\pi)^2}) \ln(x(1-x))
\nonumber\\&&
=-\frac{1}{3}(b_\rho b_\sigma- b^2       g_{\rho\sigma})
{(-\frac{13}{6})}
(\frac{i}{(4\pi)^2})
\nonumber\\&&
=\frac{i}{(4\pi)^2}
{\frac{13}{18}}
(b_\rho b_\sigma- b^2       g_{\rho\sigma})
\label{bbox.16}
\end{eqnarray}
Similar result for graph 3
\begin{eqnarray}
{\prod_{\rho\sigma}^{\rm LOG}}^{_{(3)}}
=\frac{i}{(4\pi)^2}
{\frac{13}{18}}
(b_\rho b_\sigma- b^2       g_{\rho\sigma})
\label{bbox.16.1}
\end{eqnarray}
Then the total divergent-log is
\begin{eqnarray}
\sum_{j=1}^3 
{\prod_{\rho\sigma}^{\rm LOG}}^{_{(j)}}
=\frac{i}{(4\pi)^2} 
{\frac{1}{3}}(b_\rho b_\sigma- b^2       g_{\rho\sigma})
\label{bbox.17}
\end{eqnarray}
The total divergent part is then (eqs. (\ref{bbox.8}) and (\ref{bbox.17})
\begin{eqnarray}&&
{\prod_{\rho\sigma}^{\rm DIV}}
=\frac{i}{(4\pi)^2}
{\frac{1}{6} \Big(4 b_\rho b_\sigma - b^2       g_{\rho\sigma}\Big)
}
\label{bbox.17.0}
\end{eqnarray}
%
\subsection{Convergent Parts: Graph 1}
After the shift suggested by eq. (\ref{bbox.1.10}) eq. (\ref{box.1})
becomes ($\gamma_\chi$ dropped)
\begin{eqnarray}&&
{\prod_{\rho\sigma}^{\rm CONV(1)}}(p)
={-}6\int_0^1dx \int_0^x dy ~y~\int \frac{d^D q}{(2\pi)^D}
\Big[q^2+p^2y(1-y) \Big]^{-4}
\nonumber\\&&
~ Tr ~\Bigl\{\gamma_\rho(\not\!q-y\not\!p)\not \! b
(\not\!q-y\not\!p)\gamma_\sigma (\not\!q +\not\!p(1-y))
\not \! b~(\not\!q+\not\!p(1-y))
\Bigr\}
\label{cbox.1}
\end{eqnarray}
We have 6 $q_\rho q_\sigma$- integrals and one with no q's.
We use the notations
\begin{eqnarray}&&
A1=~Tr ~\Bigl\{\gamma_\rho \gamma_\mu \not \! b
\gamma^\mu \gamma_\sigma  \not\!p
\not \! b~ \not\!p
\Bigr\}
\nonumber \\&&
=(2-D)\Big[-2p^2 b_\rho b_\sigma  
+ 2pb (b_\rho p_\sigma + p_\rho b_\sigma)
 +      g_{\rho\sigma} ( b^2 p^2- 2(pb)^2))\Big]
\nonumber \\&&
A2=  Tr ~\Bigl\{\gamma_\rho \gamma_\mu \not \! b \not\! p
\gamma_\sigma \gamma^\mu \not \! b~ \not\!p\Bigr\}
\nonumber \\&&
{
=-2b^2p^2     g_{\rho\sigma}
}
\nonumber \\&&
A3=  Tr ~\Bigl\{\gamma_\rho \gamma_\mu \not \! b \not\! p
\gamma_\sigma \not\! p  \not \! b~ \gamma^\mu \Bigr\}
\nonumber \\&&
=(2-D) \Big(4 pb  b_\rho p_\sigma - 2 p^2 b_\rho b_\sigma 
-2b^2 p_\rho p_\sigma + b^2 p^2     g_{\rho\sigma} \Big)
\nonumber \\&&
A4=
~ Tr ~\Bigl\{\gamma_\rho \not\! p \not \! b \gamma_\mu
\gamma_\sigma\gamma^\mu  \not\! b  \not \! p~ \Bigr\}
\nonumber \\&&
=(2-D)\Big(4 pb p_\rho b_\sigma - 2 p^2 b_\rho b_\sigma  - 2 b^2 p_\rho p_\sigma
+b^2 p^2      g_{\rho\sigma} \Big)
\nonumber \\&&
A5=
~ Tr ~\Bigl\{\gamma_\rho \not\! p \not \! b \gamma_\mu
\gamma_\sigma  \not\! p  \not \! b~ \gamma^\mu\Bigr\}
\nonumber \\&&
=
- 2 p^2b^2     g_{\rho\sigma}
\nonumber \\&&
A6=
 Tr ~\Bigl\{\gamma_\rho\not\!p\not \! b
\not\!p \gamma_\sigma \gamma_\mu
\not \! b~\gamma^\mu
\Bigr\}
\nonumber \\&&
=(2-D)[2pb b_\rho p_\sigma - 2 p^2 b_\rho b_\sigma 
 - 2(pb)^2      g_{\rho\sigma} +  2pb p_\rho b_\sigma 
 + p^2 b^2     g_{\rho\sigma} ]
\nonumber \\&&
A7=  Tr ~\Bigl\{\gamma_\rho\not\!p\not \! b
\not\!p\gamma_\sigma  \not\!p
\not \! b~ \not\!p \Bigr\}
\nonumber \\&&
= 8 (pb)^2p_\rho p_\sigma + 2p^4  b_\rho b_\sigma
 -4 p^2 pb(b_\rho p_\sigma + p_\rho b_\sigma)
 -b^2 p^4      g_{\rho\sigma}
\label{cbox.0.0}
\end{eqnarray}

Integration over $q$ yields (see eq. (\ref{cbox.10} )

\begin{eqnarray}&&
{\prod_{\rho\sigma}^{\rm CONV(1)}}(p)
=
-\frac{i}{(4\pi)^2}\frac{1}{3Dp^2}
6\int_0^1dx\int_0^x dy ~y~
\frac{1}{[y(1-y)]^{3-\frac{D}{2}}}
\nonumber\\&&
\Bigg( (1-y)^2 A1
- y(1-y) \Big[A2+ A3+  A4 \Big]
%
- y(1-y) A5
+ y^2 A6 
\Bigg)
\nonumber \\&&
-6\int_0^1 dx \int_0^x dy ~y~
\frac{i}{(4\pi)^2}\frac{1}{6p^4[y(1-y)]^{2-\frac{D}{2}}}
%
y^2(1-y)^2 A7.
\nonumber \\&&
\label{cbox.2.again}
\end{eqnarray}
The integration over the Feynman parameters yields
(Section \ref{sec:appB})
\begin{eqnarray}&&
{\prod_{\rho\sigma}^{\rm CONV(1)}}(p)
=
-\frac{i}{(4\pi)^2}\frac{1}{3Dp^2}
\nonumber\\&&
\Bigg( (2 A1
-  \Big[A2+ A3+  A4 +A5\Big]
+ 2 A6 
\Bigg)
%
-\frac{i}{(4\pi)^2}\frac{1}{6p^4}
 A7.
\nonumber \\&&
\label{cbox.3.again}
\end{eqnarray}
By inserting the expressions (\ref{cbox.0.0}) into (\ref{cbox.3.again})
one gets
\begin{eqnarray}&&
{\prod_{\rho\sigma}^{\rm CONV(1)}}(p)
=
-\frac{i}{(4\pi)^2}\frac{1}{12p^2} \Bigg\{
 2(2-D)
\nonumber\\&&
\Big[-2p^2 b_\rho b_\sigma  
+ 2pb (b_\rho p_\sigma + p_\rho b_\sigma)
 +      g_{\rho\sigma} ( b^2 p^2- 2(pb)^2))\Big]
\nonumber \\&&
{
+2b^2p^2     g_{\rho\sigma}
}
\nonumber \\&&
-(2-D) \Big(4 pb  b_\rho p_\sigma - 2 p^2 b_\rho b_\sigma 
-2b^2 p_\rho p_\sigma + b^2 p^2     g_{\rho\sigma} \Big)
\nonumber \\&&
-(2-D)\Big(4 pb p_\rho b_\sigma - 2 p^2 b_\rho b_\sigma  - 2 b^2 p_\rho p_\sigma
+b^2 p^2      g_{\rho\sigma} \Big)
\nonumber \\&&
+2p^2b^2
\nonumber \\&&
+2 (2-D)\Big(2pb b_\rho p_\sigma - 2 p^2 b_\rho b_\sigma 
 - 2(pb)^2      g_{\rho\sigma} +  2pb p_\rho b_\sigma 
 + p^2 b^2     g_{\rho\sigma}\Big )
\Bigg\}
\nonumber \\&&
-\frac{i}{(4\pi)^2}\frac{1}{6p^4}
\Big(
8 (pb)^2p_\rho p_\sigma + 2p^4  b_\rho b_\sigma
\nonumber \\&&
 -4 p^2 pb(b_\rho p_\sigma + p_\rho b_\sigma)
 -b^2 p^4      g_{\rho\sigma} 
\Big)
\nonumber \\&&
=
-\frac{i}{(4\pi)^2}\frac{1}{6p^2} \Bigg\{
(-4b^2  +\frac{8}{p^2}(pb)^2)p_\rho p_\sigma 
+6p^2 b_\rho b_\sigma 
\nonumber \\&&
-8  (pb)(b_\rho p_\sigma + p_\rho b_\sigma)
+(-b^2p^2 +8 (pb)^2)     g_{\rho\sigma} \Bigg\}
\label{cbox.4.collect}
\end{eqnarray}
%

\subsection{Convergent Parts: Graph 2 and 3}
%
Now we consider the graphs where the $b_\mu$ sources are
on the same $\gamma_\rho - \gamma_\sigma$
fermion line. I.e. we use the amplitude in (\ref{box.1.1})
\begin{eqnarray}&&
\prod_{\rho\sigma}(p)^{(2)}
={-}
\int \frac{d^D q}{(2\pi)^D}~ Tr ~\Bigl\{
\frac{1}{\not \! q  +i\varepsilon} ~
\gamma_\rho
\frac{1}
{
[(\not\!q+\not\!p)+i\varepsilon]}\not \! b\gamma_\chi~ 
\frac{1}{
[ \not \!q + \not\! p+i\varepsilon]
}\not \! b\gamma_\chi~
\nonumber\\&&
\frac{1}
{ [\not\!q+ \not\! p +i\varepsilon ]}
~\gamma_\sigma
\Bigr\}.
\nonumber\\&&
\label{box.1.1p}
\end{eqnarray}
With the Feynman parameterization (\ref{bbox.1.11}) one gets
\begin{eqnarray}&&
\prod_{\rho\sigma}^{\rm CONV(2)}(p)
={-}6  \int_0^1dx ~\frac{x^2}{2}~\int \frac{d^D q}{(2\pi)^D}~
\Big[q^2+p^2x(1-x) \Big]^{-4}
\nonumber\\&&
 Tr ~\Big\{
[\not \! q -x \not\!p ] ~\gamma_\rho
[\not\!q+\not\!p(1-x)]\not \! b
[ \not \!q + \not\! p(1-x))]
\not \! b
\nonumber\\&&
 [\not\!q+ \not\! p(1-x)  ]
~\gamma_\sigma
\Bigr\}
\nonumber\\&&
={-}6  \int_0^1dx ~\frac{x^2}{2}~\big(\frac{i}{(4\pi)^2}\big)
  ~\Bigg\{
\frac{1}{3p^2} \frac{1}{[x(1-x)]^{3-\frac{D}{2}}}
\nonumber\\&&
\frac{1}{D}\Big(
(1-x)^2( B1+B2+ B3)
\nonumber\\&&
-x(1-x)(B4+B5+B6)
\Big)
\nonumber\\&&
+
\frac{1}{6p^4} \frac{1}{[x(1-x)]^{2-\frac{D}{2}}}
\Big(-x(1-x)^3B7
\Big)
\Bigg\}
\label{cbox.14}
\end{eqnarray}
where
\begin{eqnarray}&&
B_1 = Tr \Big\{
\gamma_\rho \gamma_\mu\not\! b \not\! p\not \! b \not\! p\gamma_\sigma \gamma^\mu  \Big\}
\nonumber\\&&
=(2-D)\Big[  2pb(b_\rho p_\sigma-p_\rho b_\sigma) 
+       g_{\rho\sigma} (2 (pb)^2 - p^2 b^2 )\Big]
\nonumber\\&&
B_2 = Tr  \Big\{
\gamma_\rho\not\!p\not \! b\gamma_\mu\not \! b \not\! p\gamma_\sigma\gamma^\mu\Big\}
\nonumber\\&&
=-4 b^2 p_\rho p_\sigma - 4 p^2 b_\rho b_\sigma
+ 4pb(p_\rho b_\sigma + b_\rho p_\sigma )
\nonumber\\&&
+     g_{\rho\sigma} (- 4 (pb)^2 +  p^2 b^2 (6-D))
\nonumber\\&&
B_3= Tr\Big\{
\gamma_\rho\not\!p\not \! b\not\! p\not \! b\gamma_\mu\gamma_\sigma\gamma^\mu
\Big\}
\nonumber\\&&
= (2-D)\Big[ 2 pb (p_\rho b_\sigma - b_\rho p_\sigma)
+       g_{\rho\sigma}(2 (pb)^2  - p^2 b^2  )\Big]
\nonumber\\&&
B_4 =
Tr\Big\{ 
\gamma_\rho\gamma_\mu\not \! b \gamma^\mu\not \! b \not\! p  ~\gamma_\sigma \not\!p \Big\}
\nonumber\\&&
=(2-D)\Big[2b^2 p_\rho p_\sigma -b^2 p^2      g_{\rho\sigma}\Big]
\nonumber\\&&
 B_5 = Tr\Big\{ 
\gamma_\rho\gamma_\mu\not \! b \not\! p\not \! b\gamma^\mu\gamma_\sigma \not\!p\Big\}
\nonumber\\&&
= (2-D)\Big[
-2 b^2 p_\rho p_\sigma  + 2 pb (p_\rho b_\sigma + b_\rho p_\sigma)
\nonumber\\&&
+      g_{\rho\sigma} (-  2 (pb)^2  + p^2 b^2 )\Big]
\nonumber\\&&
 B_6 = Tr\Big\{
\gamma_\rho\not\!p\not \! b\gamma_\mu\not \! b  \gamma^\mu\gamma_\sigma\not\!p \Big\}
\nonumber\\&&
=(2-D) \Big[2b^2 p_\rho p_\sigma 
-p^2b^2     g_{\rho\sigma}\Big]
\nonumber\\&&
B_7 =Tr\Big\{
\gamma_\rho\not\!p\not \! b\not\! p\not \! b \not\! p\gamma_\sigma\not\!p \Big\}
\nonumber\\&&
=
(8(pb)^2-2p^2 b^2) p_\rho p_\sigma
- 2(pb)p^2(b_\rho p_\sigma +p_\rho b_\sigma)
\nonumber\\&&
+     g_{\rho\sigma}[ p^4 b^2- 2 p^2 (pb)^2]
\label{cbox.14.1}
\end{eqnarray}
In eqs. (\ref{cbox.14}) and (\ref{cbox.14.1}) $\gamma_\chi$ have been
 dropped because we consider only convergent integrals.
\par
From eqs. (\ref{cbox.10}),  (\ref{cbox.14}) and (\ref{cbox.15})
we get 
\begin{eqnarray}&&
\prod_{\rho\sigma}^{\rm CONV(2)}(p) = - \frac{i}{(4\pi)^2} \frac{1}{3Dp^2}
\Bigg\{\frac{1}{2}B1
+\frac{1}{2}B2
+\frac{1}{2}B3 
\nonumber\\&&
-B4
-B5
-B6
\Bigg\}
+\frac{i}{(4\pi)^2} \frac{1}{12p^4} B7
\label{cbox.16}
\end{eqnarray}
%

%
\par\noindent
Now we evaluate eq. (\ref{cbox.16}) by using
the 7 traces (\ref{cbox.14.1})

\begin{eqnarray}&&
\prod_{\rho\sigma}^{\rm CONV(2)}(p) = - \frac{i}{(4\pi)^2} \frac{1}{3Dp^2}
\Bigg[
\nonumber\\&&
\frac{1}{2}(2-D)
\Bigg(2pb(b_\rho p_\sigma-p_\rho b_\sigma) 
+       g_{\rho\sigma} [2 (pb)^2 - p^2 b^2 ]
\Bigg)
\nonumber\\&&
+\frac{1}{2}
\Bigg(-4 b^2 p_\rho p_\sigma - 4 p^2 b_\rho b_\sigma
+ 4pb(p_\rho b_\sigma + b_\rho p_\sigma )
\nonumber\\&&
+     g_{\rho\sigma} [- 4 (pb)^2 + p^2 b^2 (6-D)]
\Bigg)
\nonumber\\&&
+\frac{1}{2}(2-D)
\Bigg(2 pb (p_\rho b_\sigma - b_\rho p_\sigma)
+       g_{\rho\sigma}[2 (pb)^2  - p^2 b^2  ]
\Bigg)
\nonumber\\&&
-(2-D)
\Bigg(2b^2 p_\rho p_\sigma -b^2 p^2      g_{\rho\sigma}
\Bigg)
\nonumber\\&&
-(2-D)
\Bigg(
-2 b^2 p_\rho p_\sigma  + 2 pb (p_\rho b_\sigma + b_\rho p_\sigma)
\nonumber\\&&
+      g_{\rho\sigma} [-  2 (pb)^2 + p^2 b^2]
\Bigg)
\nonumber\\&&
-(2-D)
\Bigg(2b^2 p_\rho p_\sigma 
-p^2b^2     g_{\rho\sigma}
\Bigg)
\Bigg]
\nonumber\\&&
+ \frac{i}{(4\pi)^2} \frac{1}{12p^4}
\Bigg[
(8(pb)^2-2p^2 b^2) p_\rho p_\sigma
- 2(pb)p^2(b_\rho p_\sigma +p_\rho b_\sigma)
\nonumber\\&&
+     g_{\rho\sigma}[ p^4 b^2- 2 p^2 (pb)^2]
\Bigg].
\label{cbox.16.again}
\end{eqnarray}

Finally
%
\begin{eqnarray}&&
\prod_{\rho\sigma}^{\rm CONV(2)}(p)
=-\frac{i}{(4\pi)^2} \frac{1}{12p^2}
Tr\Bigg\{4 b^2 p_\rho p_\sigma - \frac{8}{p^2} (pb)^2  p_\rho p_\sigma
-2 p^2 b_\rho b_\sigma  
\nonumber\\&&
+ 8 pb (p_\rho b_\sigma + b_\rho p_\sigma)
-
8(pb)^2     g_{\rho\sigma} \Bigg\}.
\label{cbox.16.2.again}
\end{eqnarray}
By adding the contribution of graph n.3 we get
\begin{eqnarray}&&
\prod_{\rho\sigma}^{\rm CONV(2)}(p)+ \prod_{\rho\sigma}^{\rm CONV(3)}(p)
\nonumber\\&&
=-\frac{i}{(4\pi)^2} \frac{1}{6p^2}
Tr\Bigg\{4 b^2 p_\rho p_\sigma - \frac{8}{p^2} (pb)^2  p_\rho p_\sigma
-2 p^2 b_\rho b_\sigma  
\nonumber\\&&
+ 8 pb (p_\rho b_\sigma + b_\rho p_\sigma)
-
8(pb)^2     g_{\rho\sigma} \Bigg\}.
\label{cbox.16.3.again}
\end{eqnarray}
From eq. (\ref{cbox.4.collect}) we get 
\begin{eqnarray}&&  
\sum_{j=1,2,3}\prod_{\rho\sigma}^{\rm conv(j)}(p)
=
-\frac{i}{(4\pi)^2}
\frac{1}{6}
\Big[ 4  b_\rho b_\sigma - b^2     g_{\rho\sigma}
\Big].
\label{cbox.17.1pp.again}
\end{eqnarray}
Remember the POLE parts of graphs $(1)$, $(2)$ and $(3)$ from
eq. (\ref{bbox.8}) and the LOG part in eq. (\ref{bbox.17})
\begin{eqnarray}&&
\prod_{\rho\sigma}^{\rm POLE}(p)  = -\frac{i}{(4\pi)^2}  
\frac{1}{6}
\Big(-2b_\rho b_\sigma - b^2     g_{\rho\sigma}
\Big)
\label{bbox.8p}
\end{eqnarray}
\begin{eqnarray}
\sum_{j=1}^3 
{\prod_{\rho\sigma}^{\rm LOG}}^{_{(j)}}(p)
=
-\frac{i}{(4\pi)^2} 
{\frac{1}{6}}
(-2b_\rho b_\sigma +2b^2       g_{\rho\sigma}).
\label{bbox.17p}
\end{eqnarray}
The final result based on eqs. (\ref{cbox.17.1pp.again}), (\ref{bbox.8p}) 
and (\ref{bbox.17p}) is then
\begin{eqnarray}&&
\prod_{\rho\sigma}(p)
=0.
\label{cbox.18p}
\end{eqnarray}
The result in eq. (\ref{cbox.18p}) indicates that
the vacuum described by the $b_\mu$ source in the action
(\ref{CPT.02}) has no effect on massless QED, beside the
birefringence given by the Chern-Simons (\ref{CPT.01})
of the ABJ anomaly.
\par
  { 
We make explicit the above argument. By considering the
chiral transformation
\begin{eqnarray}
\psi \to e^{i \alpha \gamma_5} \psi
\label{cbox.21}
\end{eqnarray}
we get a Ward identity on the vertex functional 
generated from the path integral functional of eq.
(\ref{zerom.1})
\begin{eqnarray}
\partial_\mu \frac{\delta \Gamma}{\delta b_\mu(x)}
-i \frac{\delta\Gamma}{\delta\psi}\gamma_5\psi
+i \bar \psi \gamma_5 \frac{\delta\Gamma}{\delta\bar\psi}
=\Delta(x)\cdot \Gamma
\label{cbox.22}
\end{eqnarray}
where $\Delta\cdot\Gamma$ is the insertion of the ABJ anomaly
\begin{eqnarray}
\Delta(x)=
\frac{e^2}{(4\pi)^2}\varepsilon_{\mu\nu\rho\sigma}\partial^\mu A^\rho
\partial^\nu A^\sigma.
\label{cbox.23}
\end{eqnarray}
From a perturbative point of view, the insertion $\Delta$ brings
a $e^2$ factor in the $e$ series expansion.

Now we can differentiate eq. (\ref{cbox.22}) with respect to
$b_\nu(y), A_\rho(v), A_\sigma(w)$ and fix the external momenta
$(-p-k,0,p,k)$. With short hand notation eq. (\ref{cbox.22})
becomes
\begin{eqnarray}
(p+k)^\mu \Gamma_{\mu\nu\rho\sigma}(-p-k,0,p,k) = 
{\cal O}(e^6)
\label{cbox.24}
\end{eqnarray}
i.e.
\begin{eqnarray}
(p+k)^\mu \Gamma_{\mu\nu\rho\sigma}(-p-k,0,p,k)\Big |_{\rm one-loop} = 0,
\label{cbox.24p}
\end{eqnarray}
since the LHE of eq. (\ref{cbox.24p}) is ${\cal O}(e^2)$ and therefore
a finite value could not match the RHS.
\par
We now differentiate the above equation with respect to $p_\tau$
and then put $p=-k$. Thus finally we get at \underline{one loop}
\begin{eqnarray}
 \Gamma^{(1)}_{\mu\nu\rho\sigma}(-p-k,0,p,k)\Big |_{p=-k} =0,
\label{cbox.25}
\end{eqnarray}
i.e. the result of our explicit calculation in eq. (\ref{cbox.18p}).
\par 
Our result (\ref{cbox.18p}) disagrees with the conclusions of 
a similar investigation in Refs. \cite{Altschul:2003ce} and
\cite{Altschul:2004gs}.

} 

\normalsize

\section{Conclusions}
By using a new approach to $\gamma_5$ problem in dimensional
regularization we computed  the Chern-Simons  {effective action term}
in presence of a local CPT- and Lorentz- violating term in the Lagrangian.
No ambiguities are present in the calculation. The calculations
are straightforward  application of the algebra developed in 
Refs. \cite{Ferrari:2014jqa} and \cite{Ferrari:2015mha}.
\par
We have analyzed the different limits $m^2=0$ and $p^2=0$.
For $m^2=0$ the CPT-violating action-term can be removed
by a simple chiral transformation. The anomaly in the axial
current is the only relic of the perturbative expansion in powers
of $b_\mu$.   {The result is in agreement with the approach
based on Pauli-Villars regularization. However we understand that
other regularization procedures might yield  different
results \cite{Jackiw:1999qq}.}
\par
We  obtain that the  Chern-Simons term
is zero at one loop level for $m^2\not =0, p^2=0$
in the dimensional regularization scheme. Thus we 
 confirm the result of Pauli-Villars regularization.
\par
For virtual photons ($p^2<<m^2$) the two-point function gets a finite contribution
by the presence of the CPT- and Lorentz- violating term.
The coefficient of the $p^2$ is in agreement with previous gauge invariant
calculations (via Pauli-Villars). 
{Some} authors find a non-zero limit for $p^2=0$, i.e.
a finite Chern-Simons term (\ref{CPT.01}). This result is obtained
by abandoning dimensional regularization as in Refs.
\cite{Jackiw:1999yp}, \cite{Jackiw:1999qq},\cite{Chung:1998jv}.
\par
  {
We have devoted a consistent part of the paper to the explicit
calculation of the photon self-energy at second order
in the expansion in $b_\mu$ for the case $m^2=0$. The interest
lies in the question whether, by the removal of the  
CPT- and Lorentz- violating term,  some other relics remain beside
the Chern-Simons term. The places where to look for, are the
superficially divergent amplitudes as the box-diagram  in the photon
self-energy. After an excruciating calculation we found zero:
a result expected if ABJ anomaly is the only terms
present in the Ward identity (\ref{cbox.22}) associated to local chiral transformations.
}
\section*{Acknowledgements}
%
One of us (RF)  gratefully thanks Roman Jackiw for 
suggesting him the present subject and for challenging 
discussions. The warm and stimulating hospitality of CTP-MIT
is gratefully acknowledged. Part of this work has been
done at the Department of Physics of the University of Pisa
and of the INFN, Sezione di Pisa, to which he is very grateful.

\appendix

\section{Collection of Standard Integrals}
\label{sec:bbox}
In order to improve the presentation of the paper
we remind some Feynman formula
\begin{eqnarray}&&
\frac{1}{A_1\dots A_n} =(n-1)!\int_0^1du_1\int_0^{u_1}du_2\dots
\int_0^{u_{n-2}}du_{n-1}
\nonumber\\&&
\Big[A_1 + u_1(A_2-A_1)+\dots+u_{n-1}(A_n-A_{n-1})\Big]^{-n}
\label{bbox.1}
\end{eqnarray}

From eqs. (\ref{box.1}) and  (\ref{box.1.1}) we have the denominators
\begin{eqnarray}&&
\frac{1}{q^2q^2(q+p)^2(q+p)^2}
\nonumber\\&&
=6\int_0^1dx \int_0^x dy ~y~
\Big[(q+ y p)^2+p^2y(1-y) \Big]^{-4}
\label{bbox.1.10}
\end{eqnarray}
and 
\begin{eqnarray}&&
\frac{1}{q^2(q+p)^2(q+p)^2(q+p)^2}
\nonumber\\&&
=6\int_0^1dx ~\frac{x^2}{2}~
\Big[(q+ x p)^2+p^2x(1-x) \Big]^{-4}.
\label{bbox.1.11}
\end{eqnarray}
\par
Use Dimensional Integrals.
In Minkowski space we have
\begin{eqnarray}&&
\int \frac{d^D q}{(2\pi)^D} \frac{1}{(q^2-m^2)^k}
= i(-)^k \frac{1}{(4\pi)^\frac{D}{2}} \frac{\Gamma(k-\frac{D}{2})}{\Gamma(k)}
\frac{1}{(m^2)^{(k-\frac{D}{2})}}
\label{in.1}
\end{eqnarray}
\begin{eqnarray}&&
\int \frac{d^D q}{(2\pi)^D} \frac{q^2}{(q^2-m^2)^k}
= - i(-)^k\frac{D}{2} 
\frac{1}{(4\pi)^\frac{D}{2}} \frac{\Gamma(k-1-\frac{D}{2})}{\Gamma(k)}
\frac{1}{(m^2)^{(k-1-\frac{D}{2})}}
\label{in.2}
\end{eqnarray}
\begin{eqnarray}&&
\int \frac{d^D q}{(2\pi)^D} \frac{(q^2)^2}{(q^2-m^2)^k}
=  i(-)^k\frac{D(D+2)}{4} 
\frac{1}{(4\pi)^\frac{D}{2}} \frac{\Gamma(k-2-\frac{D}{2})}{\Gamma(k)}
\frac{1}{(m^2)^{(k-2-\frac{D}{2})}}
\label{in.3}
\end{eqnarray}
\begin{eqnarray}
\gamma = 0.5772156649
\label{form.7pp}
\end{eqnarray}
For $D\to 4$ we get ($\Gamma(z)\sim z^{-1}-\gamma+ {\cal O}(z)$)
\begin{eqnarray}&&
\int \frac{d^D q}{(2\pi)^D} \frac{q^2}{(q^2-m^2)^3}
=  \frac{i}{(4\pi)^2} \Big [ -\frac{2}{D-4}-\gamma -\frac{1}{2}
 - \ln\big(\frac{m^2}{4\pi}\big)\Big] 
\nonumber\\&&
= -\frac{i}{(4\pi)^2} \Big [ \frac{2}{D-4}+\gamma +\frac{1}{2}
 + \ln\big(\frac{m^2}{4\pi}\big)\Big] + {\cal O}(D-4)
\label{ano.5.0}
\end{eqnarray}
and 
\begin{eqnarray}
\int \frac{d^D q}{(2\pi)^D} \frac{1}{(q^2-m^2)^3}
= -\frac{i}{(4\pi)^2} 
\frac{1}{2 m^2}
\label{ano.6.0}
\end{eqnarray}
Derivative with respect to $m^2$
For $D\to 4$ we get
\begin{eqnarray}&&
\int \frac{d^D q}{(2\pi)^D} \frac{q^2}{(q^2-m^2)^4}
=  -\frac{i}{(4\pi)^2}\frac{1}{3m^2}
\label{ano.5.0.1}
\end{eqnarray}
and 
\begin{eqnarray}
\int \frac{d^D q}{(2\pi)^D} \frac{1}{(q^2-m^2)^4}
= \frac{i}{(4\pi)^2} 
\frac{1}{6 m^4}
\label{ano.6.0.1}
\end{eqnarray}
From eq. (\ref{ano.5.0})
\begin{eqnarray}&&
\int \frac{d^D q}{(2\pi)^D} \frac{q^4}{(q^2-m^2)^4}
\nonumber\\&&
= -\frac{i}{(4\pi)^2} \Big [ \frac{2}{D-4}+\gamma +\frac{5}{6}
 + \ln\big(\frac{m^2}{4\pi}\big)\Big] + {\cal O}(D-4)
\label{ano.5.0.2}
\end{eqnarray}

\section{Integrals over Feynman parameters}

\label{sec:appB}

For the log parts we need the integrals in (\ref{bbox.1.10}) and (\ref{bbox.1.11})
\begin{eqnarray}&&
6\int_0^1dx \int_0^x dy ~y~
\ln (y)= -\frac{5}{6}
\nonumber\\&&
6\int_0^1dx \int_0^x dy ~y~\ln (1-y)
=- \frac{5}{6}
\label{bbox.9}
\end{eqnarray}
Then 
\begin{eqnarray}&&
6\int_0^1dx \int_0^x dy ~y~
\ln (y(1-y))= -\frac{5}{3}.
\label{bbox.9.1}
\end{eqnarray}
Similarly
\begin{eqnarray}&&
6\int_0^1dx \frac{x^2}{2}\ln(x) = -\frac{1}{3}
\nonumber\\&&
6\int_0^1dx \frac{x^2}{2}\ln(1-x)=
{
-\frac{11}{6}
}
\label{bbox.10}
\end{eqnarray}
i.e.
\begin{eqnarray}&&
6\int_0^1dx \frac{x^2}{2}\ln(x(1-x)) = 
{- \frac{13}{6}}
\label{bbox.10.1}
\end{eqnarray}

\subsection{Integrals over Feynman parameters for $\prod^{CONV(1)}$}
We need some elementary integrals
\begin{eqnarray}&&
\int \frac{d^D q}{(2\pi)^D}
\frac{q^2}{[q^2+p^2y(1-y) ]^{4}} = \frac{i}{(4\pi)^2}\frac{1}{3p^2y(1-y)}
\nonumber \\&&
\int \frac{d^D q}{(2\pi)^D}
\Big[q^2+p^2y(1-y) \Big]^{-4} = \frac{i}{(4\pi)^2}\frac{1}{6p^4[y(1-y)]^2}
\nonumber \\&&
6\int_0^1dx \int_0^x dy ~=3
\nonumber \\&&
6\int_0^1dx \int_0^x dy ~y~=1
\nonumber \\&&
6\int_0^1dx \int_0^x dy ~\frac{y^2}{1-y}
=2
\label{cbox.10}
\end{eqnarray}
\subsection{Integrals over Feynman parameters: $\prod^{CONV(2)}$}
\par
Integrals:
\begin{eqnarray}&&
6\int_0^1 dx \frac{x^2}{2} = 1
\nonumber\\&&
6\int_0^1 dx \frac{x^2}{2} \frac{1-x}{x}=\frac{1}{2}
\label{cbox.15}
\end{eqnarray}
%
\section{Chern-Simons at generic $p^2, m^2$}
\label{sec:gen}
This Appendix includes the computation of the photon polarization {tensor} 
at first order in the parameter $b_{\mu}$ for every values of the momentum 
in the massive case. We follow the algebra of Section \ref{sec:ano}. 
We have two graphs and we consider them one by one. 
\begin{eqnarray}
\Gamma_{\rho\sigma}=-i\int \frac{d^Dq}{(2\pi)^D}\frac{Tr({\not \!   q}+m)\gamma_{\alpha}\gamma_{\chi}( {\not \!   q}+m)\gamma_{\rho } (\not \!   {q}+\not \!   {p}+m)\gamma_{ \sigma }) }{(q^2-m^2)^2((q+p)^2+m^2)}.
\label{gen.1}
\end{eqnarray} 
By introducing the Feynman parametrization and performing the usual translation
on the integration variables we easily arrive at
\begin{eqnarray}&&
\Gamma_{\mu\rho\sigma}^{\rm DIV}(k,p)=
{-i}
 \frac{1}{2}
\int_0^1  dx x
Tr ~\Bigl\{\gamma_\mu\gamma_\chi
\gamma_\alpha\gamma_\rho 
\gamma_\beta \gamma_\sigma\gamma_\iota \Bigr\}
\nonumber\\&& 
 \Bigl((1-x)p^\iota g^{\alpha\beta}
 +(-)xp^\beta g^{\alpha\iota}
+(1-x)p^\alpha g^{\beta\iota}\Bigr)
\nonumber\\&&
\Big(-\frac{i}{(4\pi)^2} \Big)\Big[\frac{2}{D-4}+\gamma +2
-\ln 4\pi+\ln(\Delta) \Big]
\nonumber\\&&
=
{-i}
 \frac{1}{2}
\int_0^1  dx x  Tr ~\Bigl\{\gamma_\mu\gamma_\chi
\gamma_\rho \gamma_\sigma\not\!p \Bigr\}
 \Bigl(2(1-x) (2-D)
 +x(6-D)\Bigr)
\nonumber\\&&
\Big(-\frac{i}{(4\pi)^2} \Big)\Big[\frac{2}{D-4}+\gamma +2
-\ln 4\pi+\ln(\Delta) \Big],
\label{gen.2}
\end{eqnarray}
where the algebra of the gamma's and $\gamma_\chi$ is performed
according to the rules of Section \ref{sec:DR}.
We separately integrate over $x$ the $x$-independent part
and  $\ln \Delta(x)$
\begin{eqnarray}&&
\Gamma_{\mu\rho\sigma}^{\rm DIV}(k,p)
=
{-i}
 \frac{1}{2}
  Tr ~\Bigl\{\gamma_\mu\gamma_\chi
\gamma_\rho \gamma_\sigma\not\!p \Bigr\}
\frac{2}{3} \Bigl(4-D \Bigr)
\nonumber\\&&
\Big(-\frac{i}{(4\pi)^2} \Big)\Big[\frac{2}{D-4}+\gamma +2
-\ln 4\pi \Big]
\nonumber\\&&
{-i}
 \frac{1}{2}\int_0^1  dx 
 Tr ~\Bigl\{\gamma_\mu\gamma_\chi
\gamma_\rho \gamma_\sigma\not\!p \Bigr\}
 \Bigl( -4x + 6 x^2
  \Bigr)
\nonumber\\&&
\Big(-\frac{i}{(4\pi)^2} \Big)
\ln(m^2-p^2 x(1-x)) 
\label{gen.2.0}
\end{eqnarray}
We use the integrals by parts
\begin{eqnarray}&&
\int_0^1 dx x \ln(m^2-p^2 x(1-x))
= \frac{1}{2}\int_0^1 dx x^2\frac{p^2(1-2x)}{m^2-p^2 x(1-x)}
\nonumber\\&&
\int_0^1 dx x^2 \ln(m^2-p^2 x(1-x))
= \frac{1}{3}\int_0^1 dx x^3\frac{p^2(1-2x)}{m^2-p^2 x(1-x)}
\label{gen.2.1}
\end{eqnarray}
Thus we get finally
\begin{eqnarray}&&
\Gamma_{\mu\rho\sigma}^{\rm DIV}(k,p)
=
{-i}
 \frac{1}{2}
\Big(-\frac{i}{(4\pi)^2} \Big)
  Tr ~\Bigl\{\gamma_\mu\gamma_\chi
\gamma_\rho \gamma_\sigma\not\!p \Bigr\}
\Bigg(-  \frac{4}{3}
\nonumber\\&&
+\int_0^1  dx  x^2 2(-1 +x)
\frac{p^2(1-2x)}{m^2-p^2 x(1-x)}\Bigg)
\label{gen.2.2}
\end{eqnarray}
The convergent integral can be evaluated as in eq.(\ref{conv.1})
with $D=4$.

\begin{eqnarray}&&
\Gamma_{\mu\rho\sigma}^{\rm CONV}(p) =
{-i}
 2
 \frac{1}{(4\pi)^2} 
\int_0^1 dx x
\frac{1}{m^2-p^2(x-x^2)}
\nonumber\\&&
Tr ~\Bigl\{\gamma_\mu\gamma_\chi~
[(1-x)\not\! p + m]  
~\gamma_\rho ~[-x\not\!p +m]\gamma_\sigma~
\nonumber\\&& 
[(1-x)\not\! p + m]
\Bigr\}
\nonumber\\&&
=
-
 \frac{1}{(4\pi)^2} \int_0^1 dx x
\frac{1}{m^2-p^2(x-x^2)}
\nonumber\\&&
\Bigg(
Tr ~\Bigl\{\gamma_\mu\gamma_\chi~ (1-x) \not\!p
~\gamma_\rho ~(-x) \not\!p\gamma_\sigma~ (1-x) \not\!p \Bigr\}
\nonumber\\&&
 + m^2[2(1-x) +x]Tr ~\Bigl\{\gamma_\mu\gamma_\chi 
 \gamma_\rho   \gamma_\sigma~
\not\!p \Bigr\} \Bigg)
\nonumber\\&&
=
-
 \frac{1}{(4\pi)^2} \int_0^1 dx x
\frac{p^2 x(1-x)^2 + m^2(2 -x)}{m^2-p^2(x-x^2)}
\nonumber\\&&
Tr ~\Bigl\{\gamma_\mu\gamma_\chi 
 \gamma_\rho   \gamma_\sigma~\not\!p \Bigr\}
.
\label{gen.2.3}
\end{eqnarray}
Now we add the expressions in (\ref{gen.2.2}) and (\ref{gen.2.3})
and introduce a factor 2 for the crossed graph.
\begin{eqnarray}&&
\Gamma_{\mu\rho\sigma}(p) =
-\frac{2}{(4\pi)^2}Tr ~\Bigl\{\gamma_\mu\gamma_\chi 
 \gamma_\rho   \gamma_\sigma~\not\!p \Bigr\} 
\Bigg( -\frac{2}{3}
\nonumber\\&&
 +
\int_0^1 dx
\frac{1}{m^2-p^2(x-x^2)} 
\nonumber\\&&
\Big[ p^2 x^2(1-x)^2 + m^2x(2 -x)  +p^2x^2(-1+x)(1-2x)
\Big]\Bigg)
\nonumber\\&&
=
-\frac{2}{(4\pi)^2}Tr ~\Bigl\{\gamma_\mu\gamma_\chi 
 \gamma_\rho   \gamma_\sigma~\not\!p \Bigr\} 
\Bigg( -\frac{2}{3}
 +
\int_0^1 dx
\frac{1}{m^2-p^2(x-x^2)} 
\nonumber\\&&
\Big[ m^2(2x -x^2)  +p^2x^2(-x^2+x)\Big]\Bigg)
\nonumber\\&&
=
-\frac{2}{(4\pi)^2}Tr ~\Bigl\{\gamma_\mu\gamma_\chi 
 \gamma_\rho   \gamma_\sigma~\not\!p \Bigr\} 
\Bigg( -1
\nonumber\\&&
 +
m^2\int_0^1 dx
\frac{2x}{m^2-p^2(x-x^2)} \Bigg).
\label{gen.2.4}
\end{eqnarray}
We use the identity 
\begin{eqnarray}
\int_0^1 dx
\frac{-1 + 2x}{m^2-p^2(x-x^2)} =
\int_0^1 dx \frac{1}{p^2}\frac{d}{dx}
\ln(m^2-p^2(x-x^2) =0.
\label{gen.2.5}
\end{eqnarray}
Thus eq. (\ref{gen.2.4}) can be written
\begin{eqnarray}&&
\Gamma_{\mu\rho\sigma}(p) =
-\frac{{2}}{(4\pi)^2}Tr ~\Bigl\{\gamma_\mu\gamma_\chi 
 \gamma_\rho   \gamma_\sigma~\not\!p \Bigr\} 
\Bigg( -1
 +
m^2\int_0^1 dx
\frac{1}{m^2-p^2(x-x^2)} \Bigg).
\nonumber\\&&
\label{gen.2.6}
\end{eqnarray}
For $p^2<<m^2$ we have
\begin{eqnarray}&&
\Gamma_{\mu\rho\sigma}(p) =
-\frac{{2}}{(4\pi)^2}Tr ~\Bigl\{\gamma_\mu\gamma_\chi 
 \gamma_\rho   \gamma_\sigma~\not\!p \Bigr\} 
\frac{1}{6} \frac{p^2}{m^2}
\label{gen.2.7}
\end{eqnarray}
which is in agreement with eq. (\ref{virtual.10}).
\par
For $p^2>> m^2$ we must restore Feynman path in 
\begin{eqnarray}&&
\Gamma_{\mu\rho\sigma}(p) =
-\frac{{2}}{(4\pi)^2}Tr ~\Bigl\{\gamma_\mu\gamma_\chi 
 \gamma_\rho   \gamma_\sigma~\not\!p \Bigr\} 
\Bigg( -1
 +
m^2\int_0^1 dx
\frac{1}{m^2-i\epsilon-p^2(x-x^2)} \Bigg).
\nonumber\\&&
\label{gen.2.6}
\end{eqnarray}
In the limit $m^2\to 0$ the factor in front 
kills the logarithmic behavior of the integral. 
Thus finally
\begin{eqnarray}&&
\lim_{m^2=0}\Gamma_{\mu\rho\sigma}(p) =
\frac{{2}}{(4\pi)^2}Tr ~\Bigl\{\gamma_\mu\gamma_\chi 
 \gamma_\rho   \gamma_\sigma~\not\!p \Bigr\} 
\label{gen.2.6}
\end{eqnarray}
as in eq. (\ref{total.1}).

\normalsize

\bibliography{reference}

\end{document}